\documentclass[12pt]{article}
\usepackage{color,xcolor,ucs}
 \usepackage{amssymb} 
  \usepackage{natbib}
\usepackage{fancyvrb} % for verbatim spacing 
\usepackage[english]{babel}
\usepackage[utf8]{inputenc}
\usepackage{amsmath}
\usepackage{amsfonts}
\usepackage{pdfpages}
\usepackage{graphicx, wrapfig}
 \usepackage{multirow}
\usepackage{booktabs}
\usepackage{floatrow}
  \usepackage{bbm}
\usepackage{amsmath}
\usepackage{algorithm}
\usepackage[noend]{algpseudocode}
\usepackage{setspace}
\makeatletter
\def\BState{\State\hskip-\ALG@thistlm}
\makeatother
\usepackage{graphicx}
\usepackage{times}

\newcommand{\bx}{\mathbf{x}}
\newcommand{\bX}{\mathbf{X}}
\newcommand{\by}{\mathbf{y}}

\newcommand{\bW}{\mathbf{W}}
\newcommand{\bTheta}{\boldsymbol{\Theta}}
\newcommand{\xopt}{\hat{\boldsymbol{\chi}}}

\newcommand{\blind}{0}

\begin{document}
	
	\def\spacingset#1{\renewcommand{\baselinestretch}%
		{#1}\small\normalsize} \spacingset{1}
	
%%%%%%%%%%%%%%%%%%%%%%%%%%%%%%%%%%%%%%%%%%%%%%%%%%%%%%%%%%%%%%%%%%%%%%%%%%%%%%

\if0\blind
{
	\title{\bf Sequential Optimization in Locally Important Dimensions}
	\author{Munir A. Winkel\\
		Department of Statistics, North Carolina State University\\
		Jonathan W. Stallings\\
		Department of Statistics, North Carolina State University\\
		Curt B. Storlie\\ 
		Division of Biomedical Statistics and Informatics, Mayo Clinic 
		and \\
		Brian Reich \\
		Department of Statistics, North Carolina State University}
	\maketitle
} \fi

\if1\blind
{
	\bigskip
	\bigskip
	\bigskip
	\begin{center}
		{\LARGE\bf Sequential Optimization in Locally Important Dimensions}
	\end{center}
	\medskip
} \fi

\bigskip	

\begin{abstract}
\noindent 

Optimizing an expensive, black-box function $f(\cdot)$ is challenging when its input space is high-dimensional.  Sequential design frameworks first model $f(\cdot)$ with a surrogate function and then optimize an acquisition function to determine input settings to evaluate next.  Optimization of both $f(\cdot)$ and the acquisition function benefit from effective dimension reduction.  Global variable selection detects and removes input variables that do not affect $f(\cdot)$ across the input space.  Further dimension reduction may be possible if we consider local variable selection around the current optimum estimate.  We develop a sequential design algorithm called Sequential Optimization in Locally Important Dimensions (SOLID) that incorporates global and local variable selection to optimize a continuous, differentiable function.  SOLID performs local variable selection by comparing the surrogate's predictions in a localized region around the estimated optimum with the $p$ alternative predictions made by removing each input variable.  The search space of the acquisition function is further restricted to focus only on the variables that are deemed locally active, leading to greater emphasis on refining the surrogate model in locally active dimensions.  A simulation study across three test functions and an application to the Sarcos robot dataset \citep{vijayakumar2000} show that SOLID outperforms conventional approaches.
\end{abstract}

\noindent%
\vfill
\hfill {\tiny technometrics tex template (do not remove)}

%\makeatletter
%\renewcommand{\Function}[2]{%
%  \csname ALG@cmd@\ALG@L @Function\endcsname{#1}{#2}%
%  \def\jayden@currentfunction{#1}%
%}
%\newcommand{\funclabel}[1]{%
%  \@bsphack
%  \protected@write\@auxout{}{%
%    \string\newlabel{#1}{{\jayden@currentfunction}{\thepage}}%
%  }%
%  \@esphack
%}
%\makeatother

\newpage
\spacingset{1.45} % DON'T change the spacing!

\section{Introduction}\label{s:introduction}
Statistical problems often involve learning about an intractable, real-valued function $f(\mathbf{x})$ that can be evaluated at given values of $p$ continuous input variables $\mathbf{x} = (x_1, ..., x_p) \in [0,1]^p$.  For example, physical experiments are often infeasible in many engineering problems so computer experiments are performed instead through evaluations of a computationally expensive simulator.  \cite{Santner2003} overview the design and analysis of computer experiments and apply the methodology to understand the evolution of wildfires \citep{berk2002}, to design a prosthesis device \citep{Chang2001}, and to optimize a helicopter blade design across 31 input variables \citep{Booker1999}.  \cite{Jala2016} recently used computer experiments to assess the impact of electromagnetic exposure on fetuses.  We focus our attention on optimization of a continuous, infinitely differentiable $f(\cdot)$, or simply $f$, that is expensive to evaluate, involves moderate to large $p$, and is measured with error.  

Due to the assumed cost of evaluation, we desire an optimization strategy that requires few evaluations.  For  expensive $f$, a sequential design approach is commonly employed to find $\boldsymbol{\chi}= \arg\max_\mathbf{x} f(\mathbf{x})$.  The approach begins with the evaluation of $f$ at an initial design of input settings.  The resulting observations are modeled with a surrogate function, $\hat{f}$, often taken to be a Gaussian Process (GP) model, and $\boldsymbol{\chi}$ is estimated from $\hat{f}$.  To improve this estimation, a new design point $\mathbf{x}^*$ is chosen based on an acquisition function that assigns a numeric value to each potential design point, which is related to the point's expected ability to improve estimation of $\boldsymbol{\chi}$ if it were added to the initial design.  The function is evaluated at $\mathbf{x}^*$, and $\hat{f}$ and $\hat{\boldsymbol{\chi}}$ are updated.

%Expected improvement ($EI$) \citep{mockus1975, jones1998} and Augmented Expected improvement ($AEI$) \citep{huang2006} are two commonly employed acquisition functions for identifying $\boldsymbol{\chi}$.  Both functions take into account the uncertainty of $\hat{f}$ so that the recommended design point $\mathbf{x}^*$ tends to either be located near the current $\hat{\boldsymbol{\chi}}$ or moves to an unobserved regions of the input space.  %The $\mathbf{x}^*$ that maximizes the acquisition function is added to the design, $y(\mathbf{x}^*)$ is observed, the surrogate model is updated with this new information, and the process repeats until resources are exhausted or a stopping criterion is reached.  

The sequential design process requires estimation of two optima at each step, that for $\hat{f}$ and the acquisition function.  Although these functions are more tractable than $f$, they are still difficult to optimize in high dimensions \citep{kandasamy2015}.  Indeed, acquisition functions are often multi-modal and contain regions where both the functions and their gradients are essentially zero, which is problematic for gradient-based optimization methods \citep{lizotte2012}.  Dimension reduction techniques are commonly employed to improve performance of maximizer estimation.  \cite{regis2016} reduce the optimization space to a trust region centered at the current estimator $\xopt$.
%\begin{equation}
%\mathcal{T}_k :=
%\big[ 
%\hat{\chi}_k - r(\boldsymbol{\omega}), 
%\hat{\chi}_k + r(\boldsymbol{\omega})
%\big] 
%\cap [0,1],
%\end{equation} with $r(\boldsymbol{\omega})$ controlling the range of the box through several inputs $\boldsymbol{\omega}$.  %The width $r(\boldsymbol{\omega})$ decreases, increases, or remains the same based on the observed value of the next design point chosen.
%centered around $\hat{\boldsymbol{\chi}}$ and then optimizes using derivative-free trust region algorithms. At each sequential run, the $AEI$ search would be restricted to a trust region 
\cite{djolonga2013} assume that $f(\mathbf{x}) = g(\mathbf{A x})$ for some smooth function $g(\cdot): \mathbb{R}^q \rightarrow \mathbb{R}$ and row-orthogonal matrix $\mathbf{A} \in \mathbb{R}^{q \times p}$ with $q <p$.  Their SI-BO algorithm uses low-rank approximation techniques to identify the subspace that supports $f$ with a Bayesian bandit framework for optimization with respect to $g$.   \cite{wang2016} propose the REMBO algorithm which uses a similar dimension reduction technique but identifies maximizers within randomly generated embeddings $\mathbf{z}=\mathbf{Ax}$ where $\mathbf{A}$ is randomly generated.  %Their method performs extremely well in large dimensions, such as $p>50$, but can perform poorly in moderately-sized $p$.

A special case of the SI-BO and REMBO algorithms could require $\mathbf{Ax}$ to simply produce a selection of $q<p$ input variables, i.e. to have the algorithms remove variables from consideration.  For example, the ``importance" of each variable may be quantified through a sensitivity analysis that assesses the variability of $f$ as $\mathbf{x}$ changes over each dimension \citep{shan2010}.  If that variability is reasonably large (small), then the variable is called globally active (inactive).   To this end, \cite{linkletter2006} specify mixture priors on the GP parameters and use Monte Carlo Markov Chains (MCMC) to determine the posterior probabilities of each variable being globally active. The globally inactive variables are removed from the design and analysis, and the resulting lower-dimensional space is easier to optimize across.

Even after employing global variable selection, further dimension reduction is possible if we were to focus our attention on a localized region of the input space, similar to the idea of a trust region.  For example, local sensitivity examines the partial derivatives of $f$ evaluated at a particular input $\mathbf{x}^*$ \citep{oakley2004}, say at $\hat{\boldsymbol{\chi}}$. \cite{bai2014} propose two approaches for local variable selection. The first assumes a local linear model around some input $\mathbf{x}$, assesses variable importance using local sensitivity, and implements a penalized LASSO framework to perform local variable selection. The second approach uses a forward/backward stepwise approach to choose the set of locally active variables around $\mathbf{x}$ using local linear estimators. \cite{zhao2018} offer a generalization of the earlier algorithms and demonstrate set convergence (of the locally active variables) as well as parameter convergence.  These papers, however, do not consider using the localized information to improve estimation of $\boldsymbol{\chi}$.

%, in order to understand the behavior of a nonlinear non-parametric system.
%We assume that the design space is the entire $[0,1]^p$ unit hypercube.  

While the ultimate goal is to identify $\boldsymbol{\chi} = \arg\max_\mathbf{x} f(\mathbf{x})$, one may not know how many additional iterations are affordable nor how many are needed to meet this goal.  Therefore, we desire a sequential design approach that consistently increases $f(\hat{\boldsymbol{\chi}})$ following each sequential run.  In this paper, we develop a Bayesian sequential design framework called Sequential Optimization in Locally Important Dimensions (SOLID) that accomplishes this by performing global variable selection and localized variable selection around $\xopt$ to optimize $f$. %, which tends to balance global and local exploration. %Our method is tailored for GP models and uses local variable selection to further optimize the AEI ($AEI$) criterion. 
% We offer guidance for choosing an initial design in Section~\ref{s:initial_design}. in Section~\ref{s:locimp}
% (Section~\ref{s:GP}) . (Section~\ref{s:GVS})
In Section~\ref{s:background}, we review Bayesian estimation of a GP, Bayesian global variable selection for response surfaces, and two common acquisition functions, expected improvement ($EI$) and augmented $EI$. We introduce in Section~\ref{s:locimp} a new measure of local variable importance, based on local changes in $\hat{f}$ near $\xopt$ after perturbing the posterior GP parameters. In Section~\ref{s:sequential}, we detail the SOLID algorithm and illustrate it on a toy example. In Section~\ref{s:simulation}, we compare SOLID with standard sequential optimization methods on three test functions and in Section~\ref{s:robot}, demonstrate SOLID's effectiveness on a robotics dataset from \cite{vijayakumar2000}.  We find that SOLID provides larger values of $f(\hat{\boldsymbol{\chi}})$ in the first few evaluations of $f$, whereas standard sequential methods require more evaluations of $f$ to obtain comparable values of $f(\xopt)$. In Section \ref{s:discussion}, we discuss the advantages and disadvantages of using SOLID to sequentially optimize an expensive black-box function and propose some areas of further development. 

\section{Background}\label{s:background}

\subsection{Gaussian Process Regression }\label{s:GP}

% suggest that a constant mean function is sufficient for interpolating the response surface between observed design points; thus ;; dictates the correlation between nearby functional values and
%The correlation function determines how much information to borrow across nearby observations when making predictions.

Let $\mathbf{X}_0$ denote the $n_0 \times p$ initial design matrix whose rows are the $p$ input settings of the $n_0$ initial runs.  The success of a sequential design strongly depends on the initial design \citep{crombecq2011} and the statistical model used to make predictions.  Space-filling designs, in which the inputs are ``spread out" across the entire design space, are a popular choice for initial designs \citep{kleijnen2005} because they maximize the possibility of identifying potential regions that contain the optimum when we have no prior information about the function. In this paper, we utilize maximin LHS designs \citep{joseph2008orthogonal} because their projection properties provide useful information for performing variable selection.

Evaluating $f$ at each row of $\mathbf{X}_0$ produces a response vector, $\mathbf{y}$, from the model $Y(\mathbf{x})= f(\mathbf{x}) + \epsilon$ where $\epsilon \sim N(0,\tau^2)$.  A surrogate model, $\hat{f}$, is constructed from $\mathbf{y}$ and is used to make predictions for an arbitrary input $\mathbf{x}$. The surrogate model considered in this paper assumes that $f$ is a realization of a Gaussian Process (GP) with mean function $E[f(\mathbf{x})] = \mu(\mathbf{x})$ and covariance function $\text{Cov}[f(\mathbf{x}),f(\mathbf{x}')] = \sigma^2 K(\mathbf{x},\mathbf{x}')$ for any two inputs $\mathbf{x}$ and $\mathbf{x}'$.  Following \cite{welch1992}, we set $\mu(\mathbf{x}) \equiv \mu$ for all $\mathbf{x}$. There are numerous choices for correlation functions, including Mat\'ern, non-stationary correlation functions, and BSS-ANOVA \citep{reich2009}. Although a non-stationary correlation function could be more appropriate, they can require a large number of design points for proper estimation, which we cannot afford for our problem of interest.  Instead, we choose the squared exponential correlation function \citep{sacks1989}
\begin{equation}\label{covar}
K( \mathbf{x}, \mathbf{x'}) =
\text{exp} \left\{
- \sum_{k=1}^p \gamma_k \left( x_{k} - x'_{k} \right)^2 
\right\}\ ,\
\end{equation} 
where $\gamma_1, ..., \gamma_p \ge 0$ are the correlation range parameters. If $\gamma_k = 0 $, then varying $x_k$ across $[0,1]$ has no effect on the response. %Note that inputs are perfectly correlated with themselves, i.e. $K(\mathbf{x},\mathbf{x}) = 1$. 
%, since observations far away are not very correlated with each other. If $\gamma_k$ is large, then the response surface is sensitive to changes in $x_k$.

The covariance function for $f$ induces a covariance function for $Y(\mathbf{x})$, which includes a nugget term $\tau^2$ to account for random variation. Even for deterministic functions where $Y(\mathbf{x}) = f(\mathbf{x})$, including a nugget effect can protect against violations of model assumptions \citep{gramacy2012}. Letting $Y_i(\mathbf{x})$ denote the $i$-th observation at $\mathbf{x}$, we then have 
\begin{equation}\label{GPregular}
\text{Cov}[ Y_i(\mathbf{x}), Y_j(\mathbf{x'})] 
=
  \left\{
\begin{array}{ll}
      \sigma^2 + \tau^2 \, \, \, &\text{if } \mathbf{x} = \mathbf{x'} \text{ and } i = j \\
      \sigma^2K(\mathbf{x},\mathbf{x'})\, \, \, &\text{otherwise.}
\end{array} 
\right.
\end{equation}
Let $\mathbf{V_X}$ be the $n\times n$ covariance matrix of $\mathbf{y}$ from design matrix $\mathbf{X}$ and let $\mathbf{v(x)}$ be the $n \times 1$ vector of covariances between $\mathbf{y}$ and new observation $Y(\mathbf{x})$. The prediction for $f(\mathbf{x})$, conditional on $\mathbf{y}$, is a Gaussian random variable with mean and variance
\begin{equation}\label{GPpred}
\begin{aligned}
\hat{f}(\mathbf{x} \mid \bTheta) 
&= 
\mu + \mathbf{v(x)}^T \mathbf{V_X}^{-1}
( \mathbf{y} - \mu \mathbf{1}_n )\ ,\
\qquad
s^2(
\mathbf{x} \mid \bTheta
) &= 
\sigma^2 
- \mathbf{v(x)}^T \mathbf{V_X}^{-1}  \mathbf{v(x)}, 
% + 
% \dfrac{ (  
% 1 -  \mathbf{1}^T \mathbf{V}^{-1}  \mathbf{v_x}
% )^2 
% }
% {  \mathbf{1}^T \mathbf{V}^{-1}  \mathbf{1 }}
\end{aligned}
\end{equation} where $\bTheta$ denotes the vector of GP parameters \citep{gelman2004}.  Of course, $\bTheta$ needs to be estimated from the available data, which we do following \cite{linkletter2006} which incorporates global variable selection, described next.

\subsection{Bayesian Estimation and Global Variable Selection}\label{s:GVS}
% One way to identify globally inactive variables is through stochastic search variable selection (SSVS), introduced by \cite{george1993}. SSVS specifies mixture priors on the model parameters and uses Monte Carlo Markov Chains (MCMC) to determine whether variables are globally active or inactive. \cite{linkletter2006} extend this tool for use in GP models by placing mixture priors on the parameters of the covariance function. 
Each input variable $x_k$ influences $f$ through its corresponding range parameter $\gamma_k$ in $K(\mathbf{x},\mathbf{x}')$, where $\gamma_k = 0$ implies that the input variable is globally inactive. \cite{linkletter2006}  places positive mass on $\gamma_k=0$ through a mixture prior such that
\begin{equation}\label{e:mixture}
\begin{aligned}
\gamma_k = u_k b_k 
,\qquad 
u_k \sim \text{Gamma}(a_u,b_u) 
,\qquad 
b_k \sim \text{Bernoulli}(\theta),
\end{aligned}
\end{equation} where $u_k$ is independent of $b_k$, and $\theta \sim \text{Beta}(a_\theta ,b_\theta)$ is the probability of each variable being globally active. More details on parameter priors are available in Appendix~\ref{s:mcmc}. 

The decision to declare an input variable globally active is based on the posterior probability $\text{Pr}(b_k = 1\mid\mathbf{y}) = \text{Pr}( \gamma_k > 0 \mid \mathbf{y}) \equiv \hat{b}_k$.  Variable $k$ is declared globally inactive if $\hat{b}_k < g$ where $g \in (0,1)$ is some threshold. %Smaller values of $g$ are preferred to prevent removing a truly active variable. 
Following \cite{linkletter2006}, a  data-driven estimate of $g$ may be found by augmenting the design with one or more random inputs and setting $g$ to be the estimated probability of those variables being active.  In this paper, once variable $k$ is deemed globally inactive, the $k^{th}$ column of $\mathbf{X}$ is permanently removed from future consideration and the remaining GP parameters are re-estimated.  Henceforth, $p$ will always reference the current number of variables that are deemed globally active at the current sequential step.  %This number may change although we have found that the change in $p$ primarily occurs after the initial design stage.

Each posterior draw $\bTheta_t$, $t=1,\dots,M$, results in a new prediction surface $\hat{f}_t=\hat{f}(\cdot|\bTheta_t)$ and $\hat{\boldsymbol{\chi}}_t$, that is estimated from $\hat{f}_t$.  %We estimate $\hat{\boldsymbol{\chi}}_t$ using $\hat{f}_t$ and its partial derivatives $\frac{ \partial } {\partial \mathbf{x} }\hat{f}_t$ as inputs to the quasi-Newton optimizer, L-BFGS-B \citep{byrd1995}.  
We will also make use of the marginal prediction surface 
$\hat{f} = M^{-1}\sum_t \hat{f}_t$ and define the estimated global maximizer to be \begin{equation}\label{e:xopt}
    \xopt = \arg\max_\mathbf{x} \hat{f}\ .\
\end{equation}
Note this estimator may differ from the alternative estimator $M^{-1}\sum_t \hat{\boldsymbol{\chi}}_t$, the average of the maximizer posterior draws.
% the marginal predicted surface is averaged over $m$ posterior draws.where $\bTheta_t$ indexes the $m$ random draws from the posterior distribution. 

\subsection{Specifying the Acquisition Function}\label{s:EI}
% In a sequential design framework, an acquisition function assigns a value for each input $\mathbf{x}$ and guides the selection of the design point at which to evaluate $f(\cdot)$ next. Expected improvement (EI) \citep{mockus1975, jones1998} assigns greater values to inputs that could maximize $f(\xopt)$, by taking both the predicted value and the predicted uncertainty into consideration. 

% We take the surrogate model $\hat{f}$ and the augmented expected improvement (AEI) criterion to decide which input to evaluate next.
To determine a new design point to help identify $\boldsymbol{\chi}$, \cite{jones1998} introduced the efficient global optimization (EGO) algorithm, which balances exploring the design space and honing in on areas likely containing $\boldsymbol{\chi}$. As introduced by \cite{mockus1975}, the improvement at any $\mathbf{x}$ is $I( \mathbf{x} ) = \max
\{  f( \mathbf{x} ) - y( \mathbf{x}_{opt} ), 0 
\}$, where $\mathbf{x}_{opt}$ is the row of $\mathbf{X}$ where $y(\mathbf{x}_{opt})$ is the largest observed response in $\mathbf{y}$. Since $f$ is unknown, the EGO algorithm instead uses $\hat{f}$ to compute the expected improvement, $EI(\mathbf{x})=E[ I(\mathbf{x})]$. \cite{jones1998} show that $EI$ can be written as
\begin{equation}\label{eq:EI}
EI(\mathbf{x}) =  
s(\mathbf{x})
\left\{
Z(\mathbf{x})\Phi\left[Z(\mathbf{x}) \right]  + \phi 
\left[Z(\mathbf{x}) \right]
\right\},
\end{equation} 
where $s(\mathbf{x})=\sqrt{s^2(\mathbf{x})}$, $Z(\mathbf{x}) = 
[\hat{f}(\mathbf{x}) -  {y}(\mathbf{x}_{opt})]/s(\mathbf{x})$, and $\Phi(\cdot)$ and $\phi(\cdot)$ are the CDF and PDF of a standard normal distribution, respectively. The next input is $\mathbf{x}^* \equiv \arg \max_{\mathbf{x}} EI(\mathbf{x})$.

% , \\ \eqref{e:marginal}
% s^2(\mathbf{x}) &= m^{-1}\sum_{t=1}^m s^2(\mathbf{x} \mid \bTheta_t),
% \end{aligned}
% \end{equation}
%Figure~\ref{f:ei2} illustrates $\hat{f}(\mathbf{x})$, $s^2(\mathbf{x})$, and the EI surface for inputs $\mathbf{x}$ across the design space $[0,1]^2$. % 

% \begin{figure}
%  \centering
%  \includegraphics[width = 1 \textwidth]{Chapter-2/f_ei3.png}
%      \caption{ 
% For $N =16$ observations in $P = 2$ dimensions, the predicted surface $\hat{f}(\cdot)$ and uncertainty $s^2(\cdot)$ built with a GP are shown. Expected improvement assigns values to inputs based on both the predicted value and the associated uncertainty. Note that the $EI$ surface can be multi-modal and contain regions where $EI$ includes the lowest allowed value, zero.
%     }\label{f:ei2}
%     \end{figure}  

% In (\ref{eq:EI}) the GP model is deterministic.
%In the deterministic case, with $\tau=0$, an input already selected will never be selected again, since its $EI$ is zero. 

The EGO algorithm was built for deterministic computer simulations, where $\tau^2 = 0$. The augmented $EI$ criterion \citep{huang2006}, or $AEI$, is more appropriate for non-deterministic functions
\begin{equation}\label{aei}
\begin{aligned}
AEI(\mathbf{x}) &\equiv E
\left[
\max \{\hat{f}(\mathbf{x}_{opt}) - \hat{f}(\mathbf{x}) ,0 \}
\right] 
\left(
1 - \dfrac{ \tau}{
\sqrt{ s^2(\mathbf{x}) + \tau^2} }
\right) \ ,\
% &= 
% E[I(\mathbf{x}_{opt})]
% \left(
% 1 - \dfrac{ \tau}{
% \sqrt{ s^2(\mathbf{x^*}) + \tau^2} }
% \right) \\
\end{aligned}
\end{equation} where $\mathbf{x}_{opt} \equiv \arg\max_{\mathbf{x}_i \in \mathbf{X}} \{ 
\hat{f}(\mathbf{x}_i) - \nu  s(\mathbf{x}_i)
\}$ for a given $\nu \geq 0$.  \cite{huang2006} state that the $\mathbf{x}_{opt}$ design point is chosen to reflect the user's degree of risk aversion, where $\nu = 1$ represents a ``willingness to trade 1 unit of predicted objective value for 1 unit of the standard deviation of prediction uncertainty." See \cite{brochu2010} for a discussion of other acquisition functions for identifying $\boldsymbol{\chi}$.%,including probability of improvement.

There are numerous algorithms to optimize $AEI$. \cite{picheny2014} optimize $AEI$ through  genetic optimization with derivatives, developed by \cite{mebane2011}.  \cite{kleijnen2015} constructs a space-filling design of candidate points $\mathbf{C} \subset [0,1]^p$ and sets the next design point to be $\mathbf{x^*} = \arg\max_{\mathbf{x} \in \mathbf{C}} AEI(\mathbf{x})$.  These approaches are not immediately appealing for the problem at hand because (1) they may fail to find the true maximizer of $AEI$ due to its multi-modal nature in moderate to high dimensions and (2) they may encourage initial exploration of the design space, leading to poor initial improvement over the current $\xopt$.  Sections~\ref{s:locimp} and \ref{s:sequential} describe how we address these issues using a local variable selection algorithm and adaptive candidate sets to improve optimization of $AEI$ and $f$.

\section{Bayesian Local Variable Selection}\label{s:locimp}

%Global variable selection helps reduce the overall dimension of the search space for optimizing the surrogate function and the acquisition function.  It may be difficult, though, to screen out variables globally without having good coverage of the input space.  
Optimizing $AEI$ around the current $\hat{\boldsymbol{\chi}}$ is appealing for multiple reasons.  For one, it limits the possibility of the next design point to explore unobserved regions of the design space having high uncertainty under the surrogate model and instead encourages identification of a local optimum in an area that has been estimated to contain $\hat{\boldsymbol{\chi}}$.  Hence it is more likely to lead to an updated $\hat{\boldsymbol{\chi}}$ with a larger $f(\hat{\boldsymbol{\chi}})$ than if we chose a design point by globally optimizing $AEI$.  Another reason is that, even when a variable is determined to be globally active, and hence has $\hat{\gamma}_k>0$, it may be that the variable is not important in a localized region of interest.  Employing local variable selection can help to optimize $AEI$ and update $\hat{\boldsymbol{\chi}}$ by significantly reducing the dimensionality of the optimization problem.  This localized strategy would be especially beneficial for expensive functions with a potentially limited number of additional evaluations.

\begin{figure}[htb]
 \centering
 \includegraphics[width = 0.85 \textwidth]{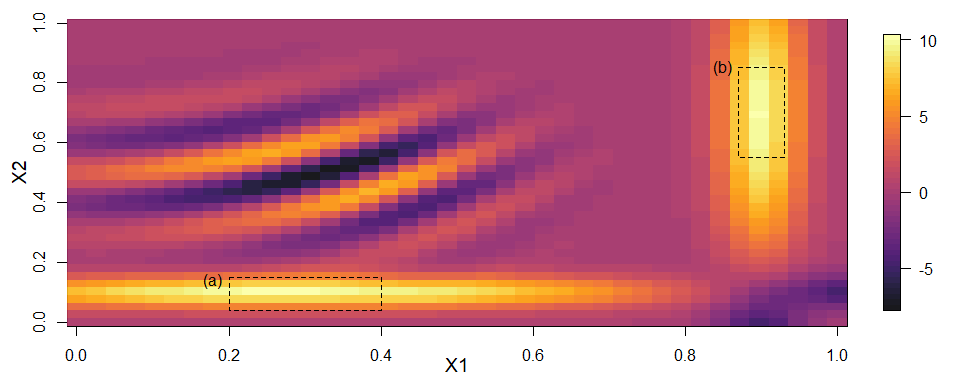}
     \caption{ 
Toy example function $f(x_1,x_2)$ in $[0,1]^2$ having two local regions in which $f$ attains its maximum. In  regions (a) and (b), optimization only needs to be done with respect to $x_2$ and $x_1$, respectively. %Similarly, in (b), optimizing $f(x_1, x_2)$ is possible by focusing only on $x_1$. 
%  Contour plot of $f(x_1, x_2)$. Both $x_1$ and $x_2$ are necessary to describe $f(x_1, x_2)$ globally and there are two local regions in which $f$ attains a global maximum. In the local region  $x_1 \in [0.85, 0.95], x_2 \in [0.6, 0.8]$, we can optimize $f(x_1, x_2)$ by focusing only on $x_1$. Similarly, in $x_1 \in [0.2, 0.3], x_2 \in [0.05, 0.15]$, optimizing $f(x_1, x_2)$ is possible by focusing only on $x_2$. 
%   Plot of $f(x_1, x_2) = (2x_2)^2 \Phi( 10x_1 - 4 ) + \sin\big( 5\pi(x_1 - x_2)\big)\Phi( 4 - 10x_1 )$, where $\Phi(\cdot)$ is the CDF of a standard normal. While $x_1$ is necessary to accurately describe $f(x_1, x_2)$ globally, once we are in a local region around the global maximizer, we can optimize $f(x_1,x_2)$ by varying only $x_2$. 
    }\label{toy}
    \end{figure}

We define here a new measure of local importance defined on some region of the input space and, in the next section, develop a flexible algorithm that uses local variable selection to identify the maximizer for $AEI$.  %find $\boldsymbol{\chi}$.
To motivate the measure, consider the two-dimensional toy example in Figure \ref{toy}. Both $x_1$ and $x_2$ are needed to describe the function globally, but there are areas that would require only one of the variables for optimization. 
% One approach to extend the global variable selection methodology in Section \ref{s:GVS} and perform local variable selection is to specify a prior on the response surface, with the probability that the response surface is zero in certain subregions \citep[e.g.][]{kang2016scalar}. However, this is likely too computationally intensive for sequential optimization.
%  \begin{wrapfigure}{r}{ 6 cm}

%
\begin{figure}
 \centering
 \includegraphics[width = 0.85 \textwidth]{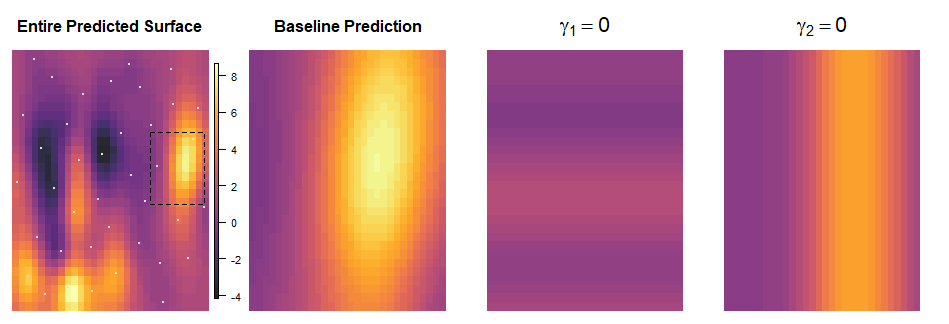}
     \caption{ 
The far left image is $\hat{f}$ for the Figure~\ref{toy} function with a highlighted region of interest, shown in greater detail in the second plot.  This represents the ``baseline" predicted surface for determining local activity.  The third and fourth plots show the alterative predicted surfaces: $\hat{f}^{1}$ with $\gamma_1 = 0$ and $\hat{f}^2$ with $\gamma_2 = 0$. The similarity between the baseline surface and $\hat{f}^{2}$ indicates that $x_2$ is locally inactive in that region. 
    }\label{f:alter}
    \end{figure}

%    
% \end{wrapfigure} 
Focusing on the localized, rectangular region labeled as (b), we make a baseline predicted surface using our current posterior estimates of $\gamma_1$ and $\gamma_2$.  The global parameter $\gamma_2$ is likely greater than 0, but $x_2$ clearly does not substantially affect $f$ in this region.  Consider the alternative predicted surface restricted to this region, where $\gamma_2$ is temporarily set to 0 and $\gamma_1$ is the same as in the baseline surface.  If this alternative predicted surface is similar to the baseline predicted surface, we would conclude that $x_2$ is locally inactive.  Figure~\ref{f:alter} shows the baseline and alternative predicted surfaces for this rectangular region and demonstrates how one would reach the conclusion that $x_2$ is locally inactive.

%, so that $x_k$ contributes nothing). 

%This is similar to the idea of total effect index \citep{homma1996, oakley2004}, where for each variable $k$, $\text{var}[E(Y| \mathbf{x}_{-k})]$ is compared with $\text{var}[E(Y)]$, where $\mathbf{x}_{-k}$ denotes all but the $k^{th}$ variable. However, our approach is both local and global. 

%  \begin{wrapfigure}{r}{ 6 cm}

%Although local variable importance can be assessed anywhere, for the purposes of finding $\boldsymbol{\chi}$, we focus our assessment to neighborhoods around $\xopt$. For each posterior draw $t \in \{1, ..., m\}$, we have a
%$\xopt_t$. 
Our approach assesses local variable importance within a neighborhood of each maximizer posterior draw $\boldsymbol{\chi}_t$ by comparing the baseline predicted surface, $\hat{f}_t$ to each of the $p$ alternative predicted surfaces, denoted by $\hat{f}_t^k$, generated by temporarily fixing $\gamma_k=0$.  To this end, we first generate $q$ prediction points $\mathbf{Q}_t$ from a truncated multivariate normal distribution 
% with mean $\hat{\boldsymbol{\chi}}_t$, variance matrix $\delta\mathbf{I}$ controlled by distance parameter $\delta$, and bounds $[0,1]^p$. %centered on that estimate of the global maximizer
\begin{equation}\label{predpoints}
\begin{aligned}
\mathbf{Q}_t \sim TN_{[0,1]^p}(\hat{\boldsymbol{\chi}}_t, \delta\mathbf{I} )\ ,\
\end{aligned}
\end{equation} 
where $\delta$ controls how far the prediction points are spread from $\xopt_t$, and the truncation keeps $\mathbf{Q}_t$ within the $[0,1]^p$ design space. We calculate the baseline and alternative predictions at the points in $\mathbf{Q}_t$, denoted $\hat{f}_t(\mathbf{Q}_t)$ and $\hat{f}_t^k(\mathbf{Q}_t)$, respectively.  We compare the baseline and alternative predictions using the squared correlation
\begin{equation}
\begin{aligned}
R^2_{kt} = \text{Corr}
\left(
\hat{f}_t(\mathbf{Q}_t), \hat{f}_t^k(\mathbf{Q}_t)\right)^2.
\end{aligned}
\end{equation} If $R^2_{kt}$ is close to one, then setting $\gamma_k = 0$ did not greatly affect the predictions, offering evidence that $x_k$ is locally inactive.

It is possible that the $\hat{\boldsymbol{\chi}}_t$'s will be dispersed across the input space and different variables are likely to be locally active with respect to different $\hat{\boldsymbol{\chi}}_t$ \citep{bai2014}. Anticipating this possibility, we average over the $R^2_{kt}$ values and define the local importance $L_k$ of input $k$ as
\begin{equation}
L_k \equiv
% 1 - \frac{1}{m}\sum_{t=1}^m R_{tk}^2.
1 - \text{mean}( R_{k1}^2, ... , R_{kM}^2).
% \left(
% R^2_{k1} , R^2_{k2}  ... , R^2_{km}  
% \rightt
\end{equation} 
Then $L_k$ is an averaged measure of local importance across the posterior distribution of $\boldsymbol{\chi}$.  We declare a variable to be locally active if $L_k \ge \rho$ for $0 < \rho < 1$ and let $\mathbf{A}$ denote the set of locally active variables. Algorithm~\ref{a:locimport} summarizes this procedure, including an additional step to perform the above calculations on only $m < M$ of the posterior draws for computational reasons.  The choice of the $m$ draws should be done carefully to be representative of the entire posterior distribution.  

%We need to best use this information about local variable importance, in order to find $\xopt$ for $f(\cdot)$. %In implementing SOLID, we choose a conservative value of $\rho = 0.02$. In our implementation, we chose a conservative value of $\rho = 0.02$. 

%Although it is possible to assess local importance around $\xopt = \arg\max_{\mathbf{x}} \dfrac{1}{m} \sum_{t =1}^m \hat{f}(\mathbf{x} \mid \bTheta_t)$, averaging across posterior draws leads to a measure of local importance that is robust to misspecification of $\boldsymbol{\chi}$.

\begin{algorithm}
\caption{Identifying Locally Active Variables}\label{a:locimport}
\begin{algorithmic}[1]
% \Procedure{Local Variable Importance}{$ \boldsymbol{\Theta}, \rho$}
		\State Initialize $\rho$ and $\delta$; randomly sample $m \leq M$ posterior draws% of the optimal design point $\hat{\boldsymbol{\chi}}_1, ... ,  \hat{\boldsymbol{\chi}}_m$
		\For{ $t \in \{1,...,m\}$}
        \State Estimate $\hat{\boldsymbol{\chi}}_t$ using $\bTheta_t$ and all globally active variables
        \State Construct $q$ prediction points $\mathbf{Q}_t$ centered around $\boldsymbol{\chi}_t$
		\State Determine baseline predictions, $\hat{f}$ at $\mathbf{Q}_t$ using $\boldsymbol{\Theta}_t$
		\For{ variable $k \in \{1,..,p\}$}
		\State Make alternative predictions, $\hat{f}^{k}_t$ at $\mathbf{Q}_t$ and calculate $R^2_{kt}$
		\EndFor
		\EndFor
		% \State If no variables meet this criteria, set $\rho = \max\{L_1, ..., L_p\}$ such that at least one variable is considered locally active.
		\State Calculate $L_k = 1-\text{mean}(R^2_{k1},...,R^2_{km})$.
		\State{\Return{ $\mathbf{A} = \{ k: L_k \ge \rho \mid \hat{\boldsymbol{\chi}} \}$ the set of locally active variables}}
% 		\EndProcedure
\end{algorithmic}
\end{algorithm} 

Declaring a variable to be locally active does not necessarily mean the function exhibits non-stationary behavior.  For a function generated from a stationary GP, the parameters $\gamma_k$ describe the correlation with respect to the entire input space.  As we focus our attention to a smaller region of interest, variables having $\gamma_k >0$ will start to appear unimportant.  The larger $\gamma_k$ is, the smaller the region needs to be for this to happen.  We allow the uncertainty of $\boldsymbol{\chi}$ to dictate the size of the region.  Even if $f$ is generated from a non-stationary GP, our use of a surrogate function assuming a stationary GP is necessary given our cost assumptions of evaluating $f$.  To further support this, our toy example and numerical studies involve functions that exhibit non-stationary behavior.

\section{Sequential Optimization using SOLID}\label{s:sequential} 

In practice, finding the optimal $AEI$ often involves reducing the optimization space, such as with trust regions.  %including a contraction parameter, expansion parameter, initial length, minimum length, maximum length, actual improvement $AI$ (as opposed to expected improvement) for the previous run, $EI$ for the previous run, and a pre-specified ratio $\eta$. $r(\omega)$ increases if ${AI}/{EI} \ge \eta$, decreases if $AI = 0$ (and a sufficient number of design points remain within the contracted region), or otherwise remains the same.
If $\hat{\boldsymbol{\chi}}$ is near $\boldsymbol{\chi}$, further exploration would be unnecessary, and restricting the search for the $AEI$ maximizer to a small neighborhood of $\hat{\boldsymbol{\chi}}$ would be advantageous.  Local variable selection can further reduce the dimension of this localized search space.  We detail here the SOLID algorithm that utilizes global and local variable importance measures to improve estimation of $\boldsymbol{\chi}$.

In SOLID, rather than restrict the $AEI$ search space for the $k$-th locally active variable to be within some neighborhood centered at $\hat{\boldsymbol{\chi}}$, we restrict the space to be
\begin{equation}\label{rspace}
\mathcal{R}^{\delta}_k := 
\big[
% \max{\{0 , 
\min{  (  \hat{\chi}_{k,1}, ...,  \hat{\chi}_{k,m} )}  - \delta\
% }} 
, 
% \min{\{1 , 
\max{  (  \hat{\chi}_{k,1}, ...,  \hat{\chi}_{k,m} )}  + \delta\
% }} 
\big]  \cap [0,1]\ ,\
\end{equation} using the $k$-th coordinate of the $m$ $\hat{\boldsymbol{\chi}}_t$'s and $\delta$ implemented in Algorithm 1.  For the $j$-th locally inactive variable, we set $\mathcal{R}^{\delta}_j := \hat{\chi}_j$, the $j^{th}$ coordinate of $\xopt$ which is calculated from $\hat{f}$.  Let $\mathcal{R}^{\delta}$ denote the corresponding restricted search space.  

Unlike the trust region in \cite{regis2016}, the range of the $\mathcal{R}^\delta$ search is guided directly by the estimates of $\hat{\chi}_{k,t}$ and explores only the locally active variables.  Moreover, \eqref{rspace} incorporates the uncertainty of $\boldsymbol{\chi}$ into the search space $\mathcal{R}^\delta$.  The inclusion of the $\delta$ parameter allows us to further expand the region if the distribution of the $\hat{\chi}_{k,1}, ...,  \hat{\chi}_{k,m}$ may be too narrow (perhaps due to selecting a smaller $m$ for better computational performance).  One could also use a different parameter than the $\delta$ used in Algorithm 1.  

To search for the $AEI$ optimum in $\mathcal{R}^\delta$, we construct an $|\mathbf{A}|$-dimensional maximin LHS design $\mathbf{L}_\delta\subset \mathcal{R}^\delta$ with $c$ settings to evaluate $AEI$.  For example, if only the first $a < p$ variables are locally active, then the set of restricted candidate points $\mathbf{C}_\delta \subseteq \mathcal{R}^\delta$ would be
% \begin{equation}
% \mathbf{C}_\mathbf{A} 
% =
% \left[
% \begin{matrix}
% L_{11} & L_{12} & ... & L_{1a} & \hat{\boldsymbol{\chi}}_{a+1} & ... & \hat{\boldsymbol{\chi}}_{p} \\
% L_{21} & L_{22} & ... & L_{2a} & \hat{\boldsymbol{\chi}}_{a+1} & ... & \hat{\boldsymbol{\chi}}_{p} \\
% \vdots  & \vdots & \vdots & \vdots & \vdots & \vdots & \vdots \\
% L_{C1} & L_{C2} & ... & L_{Ca} & \hat{\boldsymbol{\chi}}_{a+1} & ... & \hat{\boldsymbol{\chi}}_{p} \\
% \end{matrix}
% \right].
% \end{equation}
\begin{equation}\label{smallC}
\mathbf{C}_\delta
=
\left[
\begin{matrix}
\mathbf{L}_\delta
& \hat{{\chi}}_{a+1} \mathbf{1}_{c}
& ... 
& \hat{{\chi}}_{p-1} \mathbf{1}_{c} 
& \hat{{\chi}}_{p}\mathbf{1}_{c}   
\end{matrix}
\right].
\end{equation}
It is possible that $\mathcal{R}^{\delta}$ is still too restrictive so we also consider a slightly larger space $\mathcal{R}^\mathbf{A}$ with $\mathcal{R}_k^\mathbf{A} := [0,1]$ for each locally active variable $k$, and $\mathcal{R}_j^\mathbf{A} := \hat{\boldsymbol{\chi}}_j$ for each locally inactive variable $j$.  This allows us to consider exploration of unobserved locations, but only within the locally active dimension. We again use a maximin LHS design $\mathbf{L}$ of dimension $|\mathbf{A}|$, within $\mathcal{R}^\mathbf{A}$ for the $c$ candidate points.  These unrestricted candidate points $\mathbf{C}_\mathbf{A}$ are constructed in the same manner as in (\ref{smallC}), where the column of each locally inactive variable $j \in \mathbf{A}^c$ is $\hat{\chi}_{j} \mathbf{1}_{c}$.  Note that $\mathbf{C}_\delta \subseteq \mathbf{C}_\mathbf{A}$ but that $\mathbf{C}_\delta$ is more densely concentrated around the $\hat{\boldsymbol{\chi}}_t$'s.

While it is possible to combine both $\mathbf{C}_\delta$ and $\mathbf{C}_\mathbf{A}$ into one large candidate set, we have found that the candidate points with the greatest $AEI$ often all reside in one of the two sets. Whichever set has the largest $AEI$ becomes the final set of candidate points $\mathbf{C}$.  Conceptually, this helps us see if SOLID is honing-in on a restricted space $\mathcal{R}^{\delta}$ or exploring the larger space $\mathcal{R}^\mathbf{A}$.  Using the $|\mathbf{A}|$-dimensional gradient of $AEI$ (see \cite{picheny2014}), we conduct line searches from the five most promising candidates in $\mathbf{C}$, restricting the search to lie within a ball of radius $\delta$ (as specified in \eqref{predpoints}). After the line searches are complete, the one with the largest $AEI$ is chosen as the next design point. 

Thus far, we have described using the local variable selection results only for optimizing $AEI$, but they could also apply for the estimate of $\hat{\boldsymbol{\chi}}$.  One may be skeptical of doing this since the proposed local variable selection procedure uses the posterior draws $\hat{\boldsymbol{\chi}}_t$ which are calculated without local variable selection.  By estimating $\boldsymbol{\chi}$ across all globally active variables, we are likely to observe larger variation of $\boldsymbol{\chi}_t$ because we are optimizing in higher dimensions.  Modifying the $\hat{\boldsymbol{\chi}}$ optimization space to $\mathcal{R}^\mathbf{A}$ should provide a more stable estimator.

We have described, up to this point, a single iteration of a sequential optimization algorithm detailing the global and local variable selection procedure, the localized $AEI$ optimization, and the localized estimation of $\boldsymbol{\chi}$.  The next iteration starts with evaluation of the recommended design point from the localized $AEI$ optimization.  We then re-estimate the GP parameters using MCMC with priors from the initial iteration.  Recall, if at least one variable is deemed globally inactive at this next step then we would again re-estimate the GP parameters and perform global variable selection using only the globally active variables.  Next we perform our Bayesian local variable selection across all globally active variables and maximize $AEI$ in the new localized region.  Note that previous results of the local variable selection algorithm are ignored.  This way, any misclassification of locally active/inactive variables will not have long-lasting consequences.  This flexibility also allows assessment of local importance to recalibrate when $\hat{\boldsymbol{\chi}}$ changes.  The SOLID procedure is summarized in Algorithm~\ref{a:SOLID}.

\begin{algorithm}\caption{Summary of SOLID}\label{a:SOLID}
\begin{algorithmic}[1]
% 	\Procedure{ SOLID }{}
        \State Set $n_0$, $N$ (maximum number of evaluations), $g$, $\delta$, $\rho$, $M$, $m$, $c$
		\State {Create an initial maximin LHS$(n_0,p)$ design, $\mathbf{X}$}
		\State Generate $\mathbf{y}$ from $Y(\mathbf{X})$
		\For{step $i\in\{0,...,N\}$}
		\State {Obtain $M$ posterior draws of $\boldsymbol{\Theta}_t$ and $\boldsymbol{\chi}_t$ (Section~\ref{s:GVS}); calculate $\hat{f}$ and $\hat{\boldsymbol{\chi}}$.}
		\State \textbf{Global Variable Selection}: Remove variables with $\hat{b}_k < g$ from $\mathbf{X}$; if variables removed, repeat step (5) with new $\mathbf{X}$
		\State \textbf{Local Variable Selection}: Implement Algorithm~\ref{a:locimport} with $\delta$, $\rho$, and $m< M$ $\boldsymbol{\chi}_t$'s; store $\mathbf{A}$
%         using \Call{LVS}{$ \boldsymbol{\Theta}, \rho, i$}
		\State Define restricted $\mathcal{R}^\delta$ and unrestricted $\mathcal{R}^\mathbf{A}$ search spaces
		\State \textbf{Localized Optimum Estimation}: Update estimate $\hat{\boldsymbol{\chi}}$ in $\mathcal{R}^\mathbf{A}$ using $\hat{f}$; store as $\hat{\boldsymbol{\chi}}^i$.
        \State Create maximin LHS designs $\mathbf{C}_\delta \subseteq \mathcal{R}^\delta$ and $\mathbf{C}_\mathbf{A} \subseteq \mathcal{R}^\mathbf{A}$
        \State Evaluate $AEI$ in $\mathbf{C}_\delta$ and $\mathbf{C}_\mathbf{A}$; define the set with the largest $AEI$ as $\mathbf{C}$
        \State \textbf{Localized $\boldsymbol{AEI}$ Estimation}: Perform line search optimization to identify $\mathbf{x^{*}} = \arg\max_{\mathbf{x}\in \mathbf{C}} AEI(\mathbf{x})$ 
%         using \Call{Candidates}{$\mathbf{A}, \boldsymbol{\Theta}, \delta$}
% 		\State Choose the resulting $\mathbf{x^{*}}$ that maximizes $AEI$
%         using \Call{AEI}{$
% 			\mathbf{A},
% 			\mathbf{C},\boldsymbol{{\Theta}}$
%             }
		\State Augment $\mathbf{x}^{*}$ to $\mathbf{X}$; generate $Y(\mathbf{x}^{*})$ and add to $\mathbf{y}$
		%\State Re-estimate the GP parameters $\boldsymbol{\Theta}$
		\EndFor
% 		\EndProcedure
\State \Return $\{\hat{\boldsymbol{\chi}}^0,\dots,\hat{\boldsymbol{\chi}}^N\}$
\end{algorithmic}
\end{algorithm}

\begin{figure}[h]
 \centering
 \includegraphics[width = .5 \textwidth]{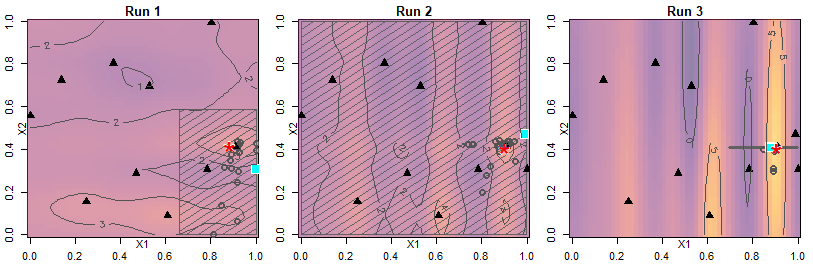}
 \includegraphics[width = .5 \textwidth]{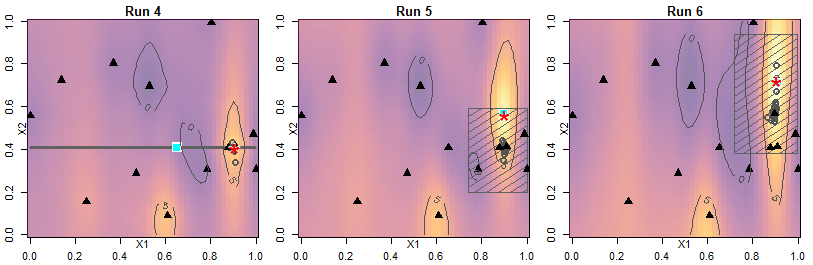}
 \includegraphics[width = .5 \textwidth]{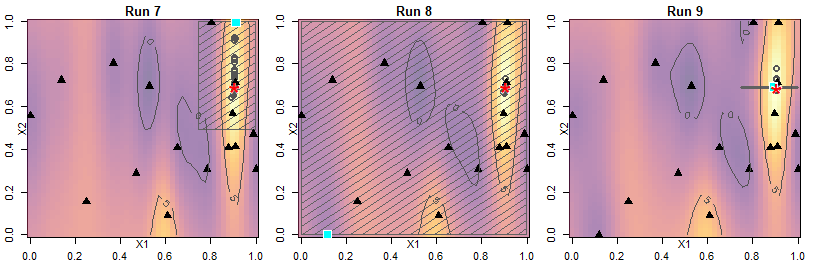}
 
     \caption{ 
     The predicted surface $\hat{f}$ of the Figure~\ref{toy} function across nine runs of SOLID. The design points ($\blacktriangle$) and $\hat{\boldsymbol{\chi}}$ ($\huge{\color{red}{*}}$) are shown. Local importance is assessed at each of the 25 posterior draws of $\hat{\boldsymbol{\chi}}_t$ ($\circledcirc$). The shaded rectangles represents either $\mathcal{R}^\mathbf{A}$ or $\mathcal{R}^\delta$, depending on which has the $AEI$ maximizer ($\color{cyan}{\blacksquare}$). In runs 3, 4, and 9, $x_2$ was found to be locally inactive, so the candidate points explore only the $x_1$ dimension (horizontal line). 
    }\label{solidtoy}
    \end{figure}

We demonstrate SOLID in Figure \ref{solidtoy} on the toy function (Figure \ref{toy}) that includes a third, unimportant variable $x_3$.  We set the global and local thresholds to be $g = 0.50$ and $\rho = 0.30$, and set $\delta = 0.15$. We considered $c = 300$ candidate points when optimizing $AEI$. Observations were generated with noise, $\tau^2 = 0.08$. For simplicity, the marginal surfaces were built using $m = 25$ random draws using MCMC chains of length $M = 500$.  We start with an initial maximin LHS design with $n_0=10, p =3$, shown as $\blacktriangle$ in Figure~\ref{solidtoy} in the upper left panel.

In the first iteration, $\mathbf{y}(\mathbf{x}_{opt})=7.71$ and all three variables were deemed globally active. %$\hat{\mathbf{b}} = (0.78, 0.91, 0.89$). 
Local importance was assessed around the $m = 25$ posterior draws of $\hat{\boldsymbol{\chi}}_t$, shown as open circles in Figure~\ref{solidtoy}. All three variables were also deemed locally active, with $L_1 = 0.52, L_2 = 0.81,$ and $L_3 = 0.76$.  It follows then that $\mathcal{R}^\mathbf{A}=[0,1]^3$, the entire input space, and $\mathcal{R}^\delta$ (visualized in just the important dimensions as the shaded rectangle in the upper left panel of Figure~\ref{solidtoy}) was bounded by $0.66 \le x_1 \le 1.00, 0.00 \le x_2 \le 0.59$ and $0.44 \le x_3 \le 0.94$.  The localized optimum estimation step was not technically localized since $\mathcal{R}^\mathbf{A}=[0,1]^3$ and we determined $\hat{\boldsymbol{\chi}}^0=(0.88, 0.42, 0.57)$.

Next we optimized $AEI$ to determine the next design point.  The $c=300$ candidate points were generated in $\mathcal{R}^\mathbf{A}$ and $\mathcal{R}^\delta$, producing $\mathbf{C}^\mathbf{A}$ and $\mathbf{C}^\delta$, respectively.  The set $\mathbf{C}_{\delta}$ contained the point with the largest $AEI$ and a line search optimization algorithm contained in this restricted space determined the next design point to be $\mathbf{x}^* = (1.00, 0.31, 0.67)$.  This point was added to $\mathbf{X}$ and we then evaluated $Y(\mathbf{x}^*)$.

%is added to $\mathbf{X}$. The GP parameters are then re-estimated, but $\hat{{\boldsymbol{\chi}}}$ remains the same at , 
%, where $y(\mathbf{x}^*) = 0.28$. 
% with $\hat{f}(\hat{{\boldsymbol{\chi}}}) = 7.60$. 
%with true value $f({\hat{\boldsymbol{\chi}}})= 6.98$. %This is lower than the largest observed $y$, in part because the $GP$ model is still assessing how noisy the observations are. 
In the second run, all variables were again found to be globally and locally active.  The updated optimum was $\hat{{\boldsymbol{\chi}}}^1=(0.90, 0.40, 0.58)$.  Here the unrestricted candidate set $\mathbf{C}_\mathbf{A}=[0,1]^3$ was preferred for $AEI$ optimization and the next selected point was $\mathbf{x}^* = (0.99, 0.47, 0.55)$, close to $\hat{{\boldsymbol{\chi}}}^1$.  %next and 
%with $y(\mathbf{x}^*) = 0.80$.
%re-estimating the GP parameters, , yielding %with $\hat{f}(\hat{{\boldsymbol{\chi}}}) = 7.29$ and 
%$f({\hat{\boldsymbol{\chi}}})= 7.39$.

% , where $y(\mathbf{x}^*) = 6.55$ The restricted candidate set is preferred, with $0.70 \le x_1 \le 1.00$ and $x_2 = 0.41$.

In the third run, we found $\hat{b}_3=0.49 < g=0.5$ so variable $x_3$ was permanently removed and the remaining GP parameters were re-estimated.  Both $x_1$ and $x_2$ were still deemed globally active, as they should be.  Their respective local importance measures were $L_1 = 0.85$ and $L_2 = 0.29$.  With $L_2 < \rho=0.3$, $x_2$ was declared locally inactive, and so $\mathcal{R}^\mathbf{A}$ and $\mathcal{R}^\delta$ were entirely contained within the $x_1$ dimension, both fixing $x_2$ at $0.41$.  Maximizing $AEI$, the next design point $\mathbf{x}^* = (0.88, 0.41)$ was added to $\mathbf{X}$.  Note that the setting for the third input variable was no longer considered.  If a value for that variable were required for the function to be evaluated, one could choose the corresponding coordinate from the previous optimum estimate, which in this case was $0.58$ for $x_3$.

%Near the now two-dimensional global maximizer, $\hat{{\boldsym bol{\chi}}} = (0.90, 0.41)$,   The GP parameters are re-estimated, and in refining the estimate of the global maximizer, $\hat{\chi}_2$ is fixed at $0.41$. Only $x_1 \in [0,1]$ is explored, leading to $\hat{{\boldsymbol{\chi}}} = (0.90, 0.41)$ with 
% $\hat{f}(\hat{{\boldsymbol{\chi}}}) = 7.48$ and 
%true value $f({\hat{\boldsymbol{\chi}}})= 7.42$. 
%(but fixing $x_2 = 0.69$)

For the remaining runs, $x_1$ and $x_2$ were always found to be globally active but $x_2$ was found locally inactive in runs 4 and 9.  As design points were added and parameters were updated, $f({\hat{\boldsymbol{\chi}}})$ rose steadily: $7.42, 9.41, 9.89, 9.92, 9.93$, and $10.00$, with $10.00$ being the largest possible value. %In the ninth run, $x_2$ is locally inactive, and a restricted candidate set is preferred. Adding $\mathbf{x}^* = (0.89, 0.69)$ 

\section{ Simulation Study }\label{s:simulation}

We conducted a simulation study to evaluate the effects of global and local variable selection on sequential optimization. We compared four approaches:
% \item {\bf None} does not do any variable selection;
(1) \textbf{GVS} conducts global variable selection only; (2) \textbf{SOLID}, as described in Section~\ref{s:sequential}; (3)
% \item {\bf Both} conducts global and local variable selection, but once a variable is declared 'locally inactive,' it will never be considered again; and
\textbf{Oracle} uses only the known globally active variables (without performing any variable selection); and (4) \textbf{None} uses all variables.  Within each simulation run, all four approaches used the same initial design, a maximin Latin Hypercube design, and the same vector of initial responses.  The responses were measured with error, $\tau^2=0.05$.

\begin{figure}[ht]
  \centering
  	\includegraphics[width=1.5in]{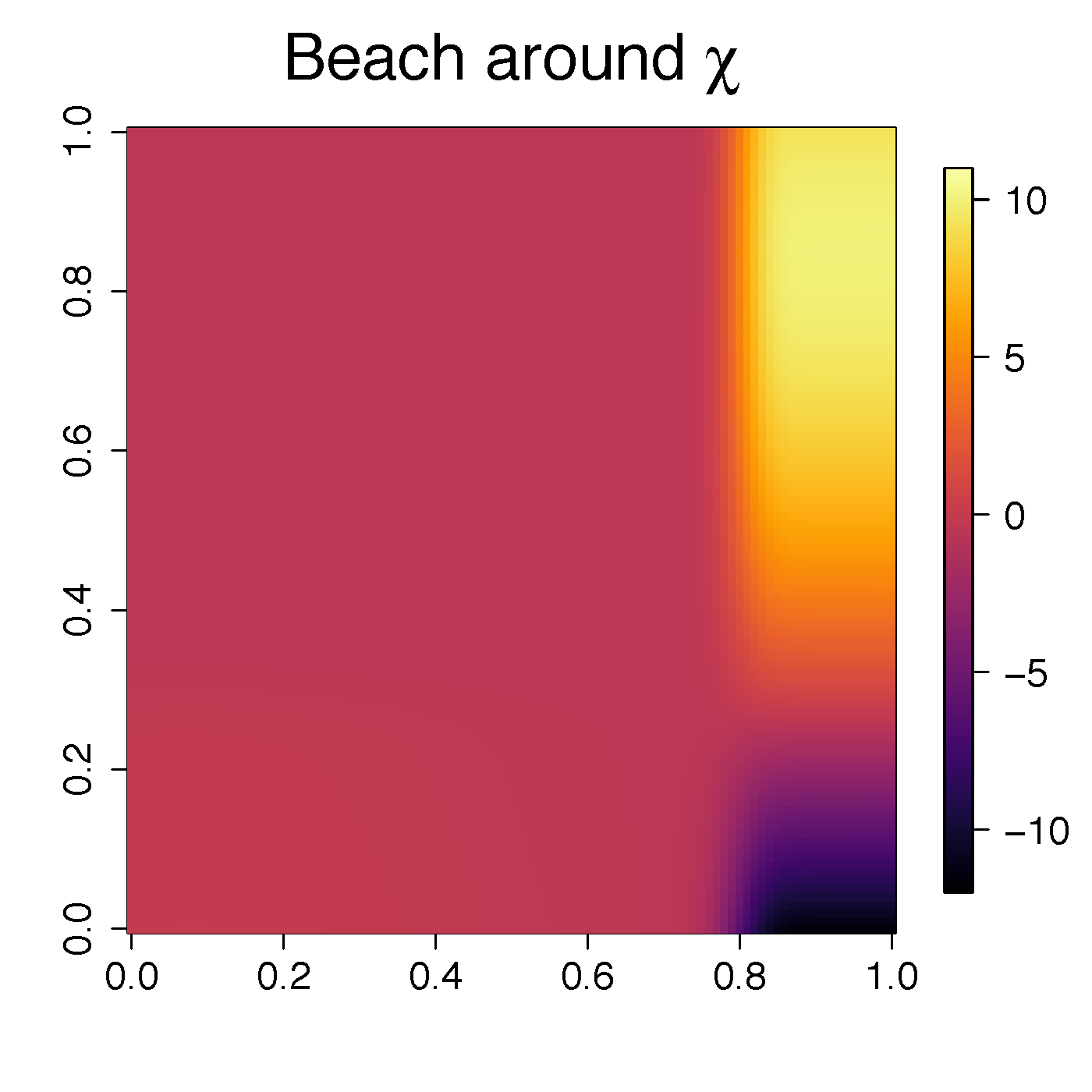} 
  	\includegraphics[width=1.5in]{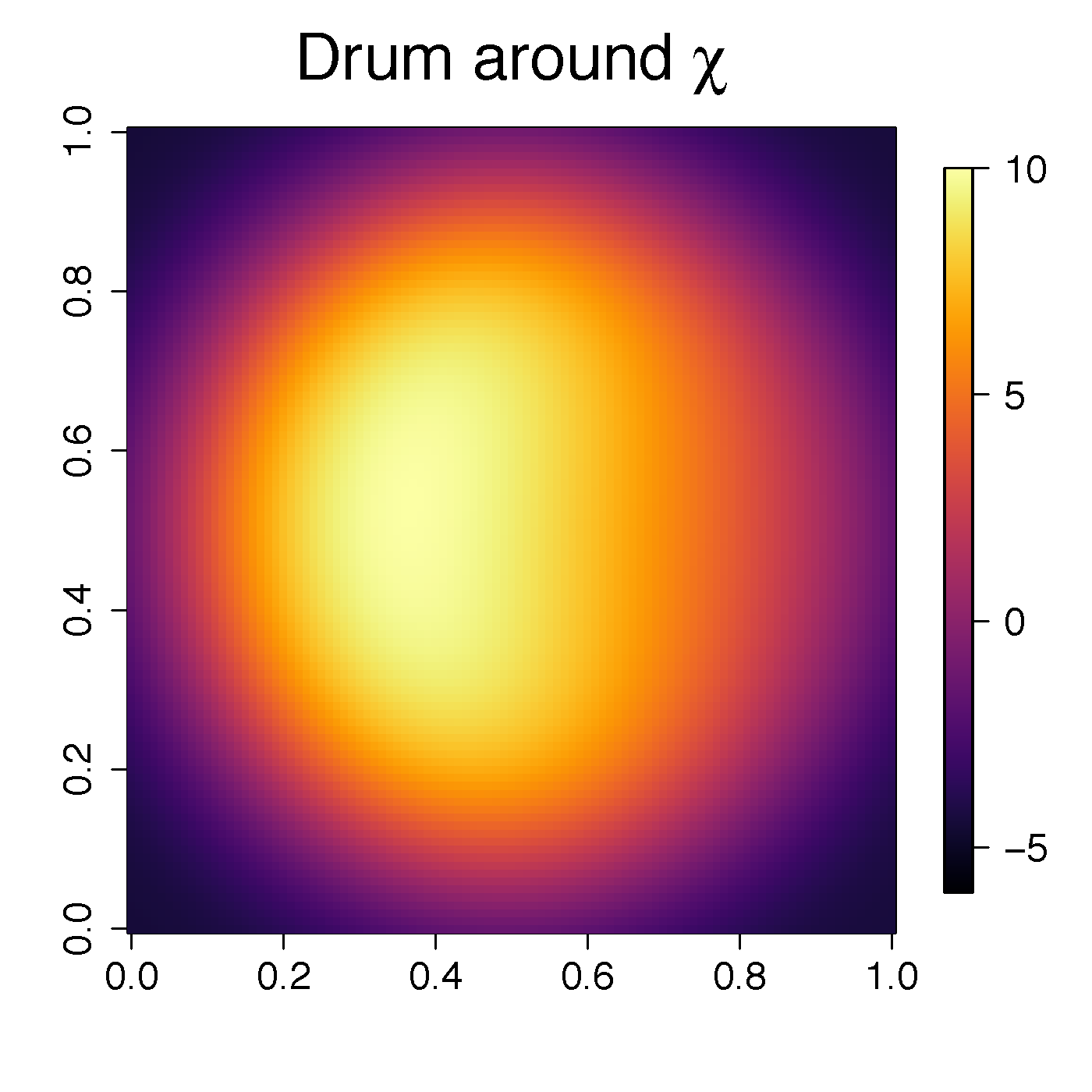} 
	\includegraphics[width=1.5in]{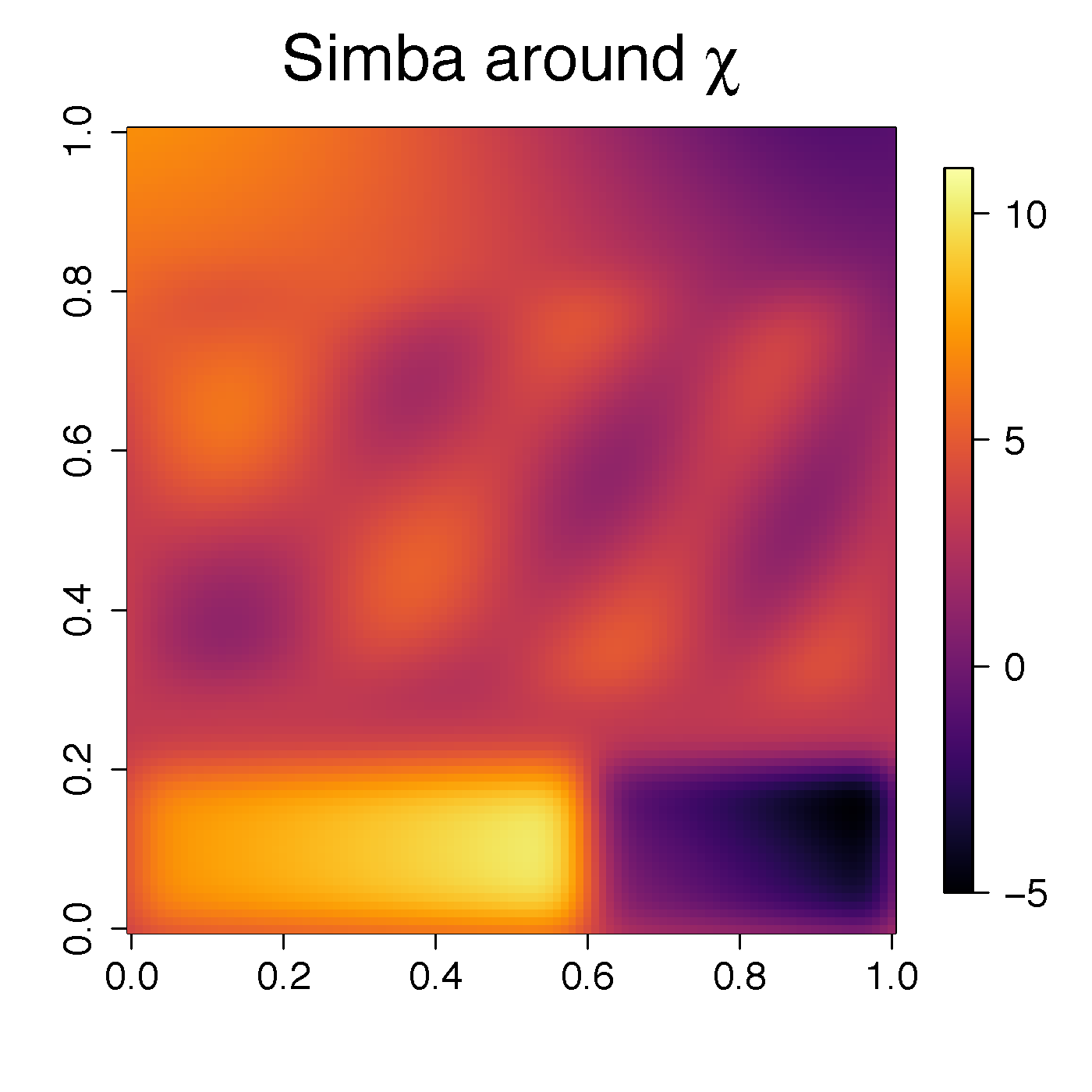}\\ 
	\includegraphics[width=1.5in]{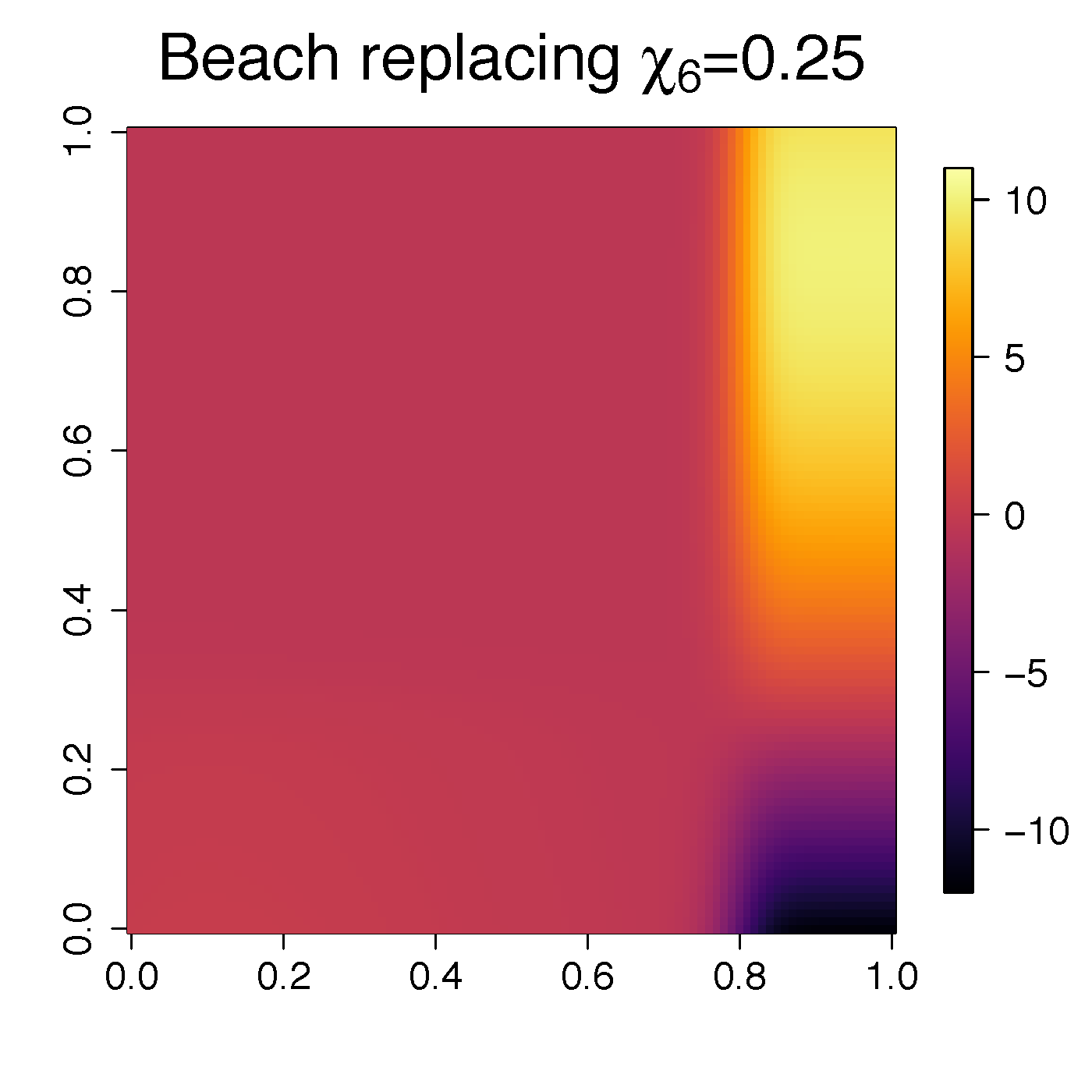} 
	\includegraphics[width=1.5in]{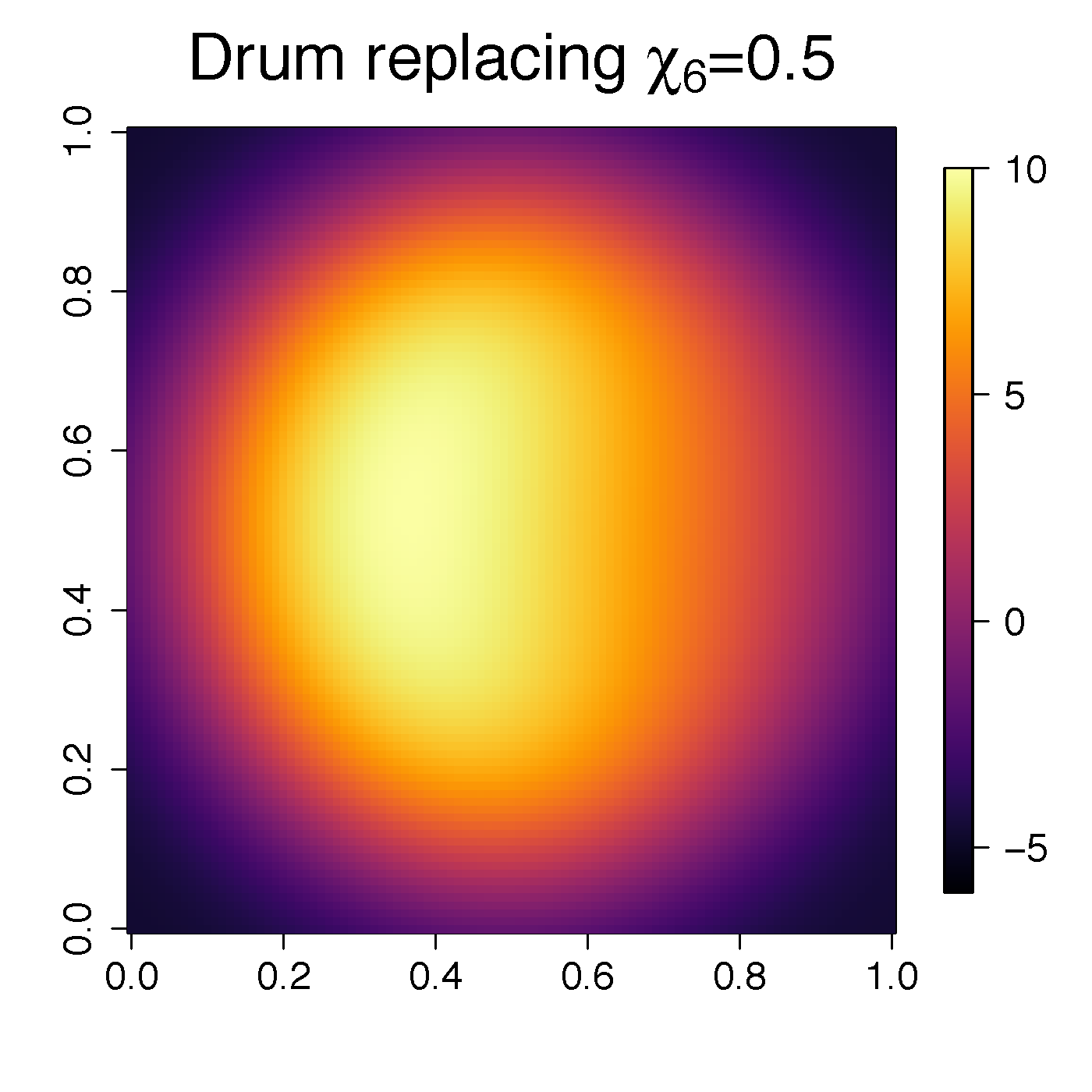} 
	\includegraphics[width=1.5in]{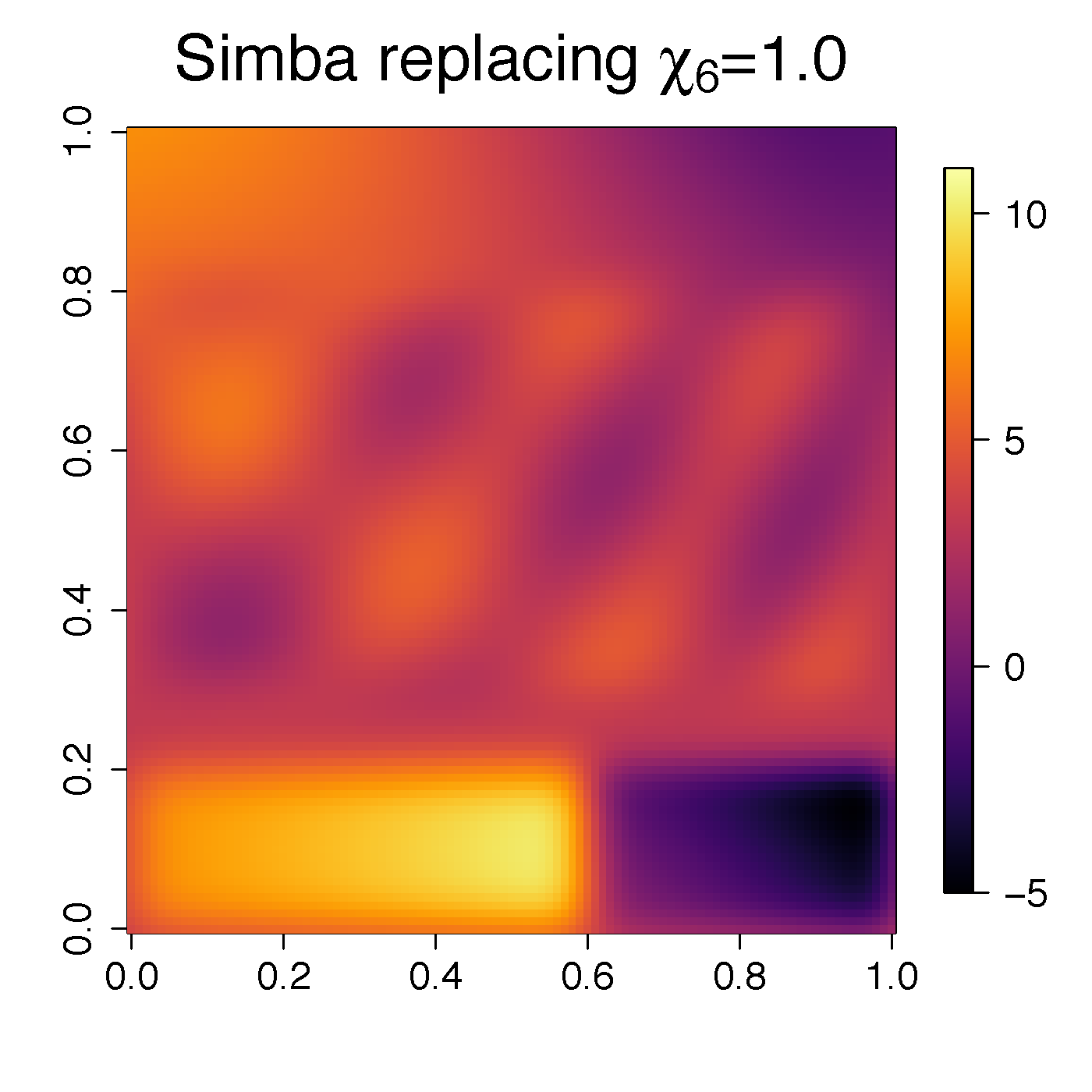}\\ 
	\includegraphics[width=1.5in]{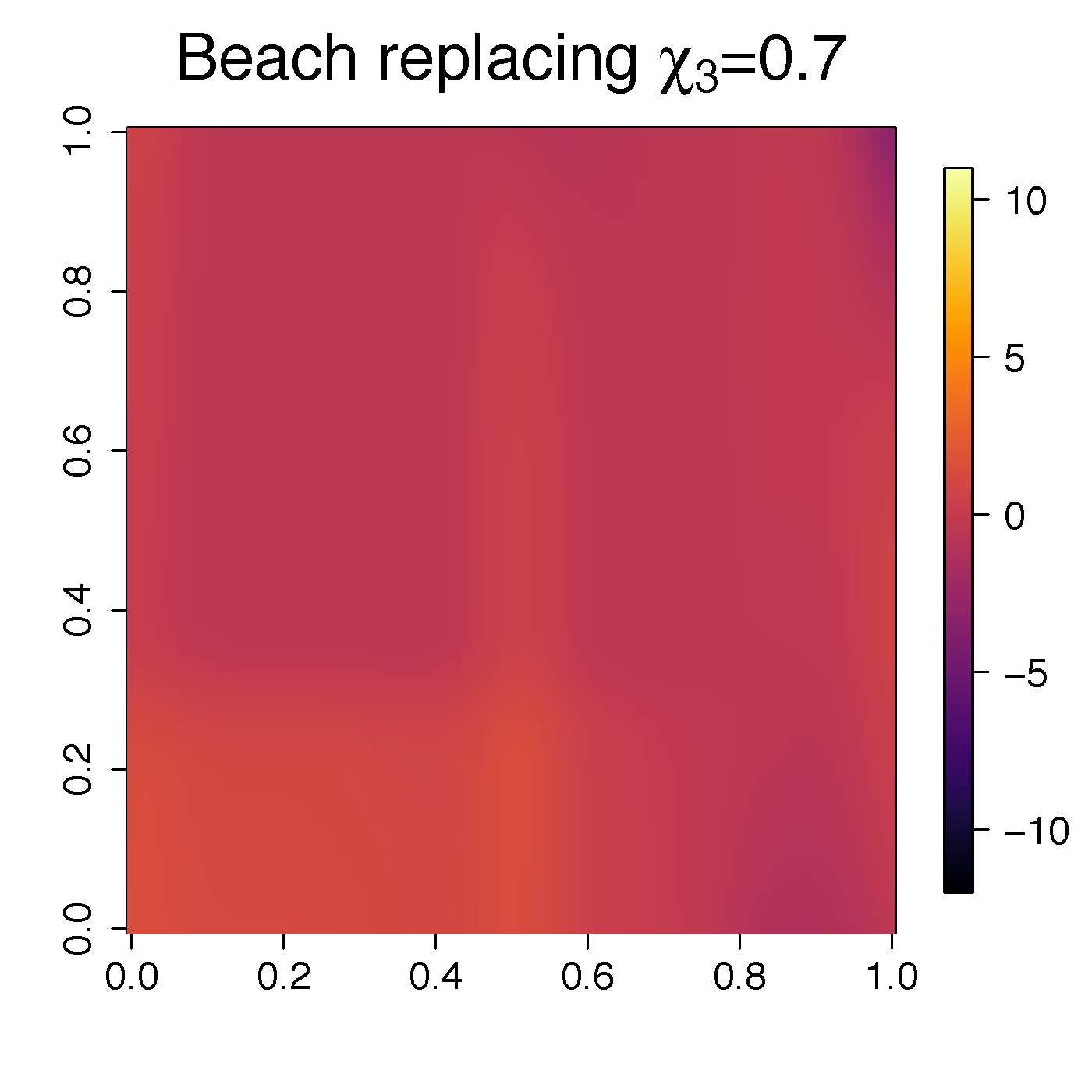} 
	\includegraphics[width=1.5in]{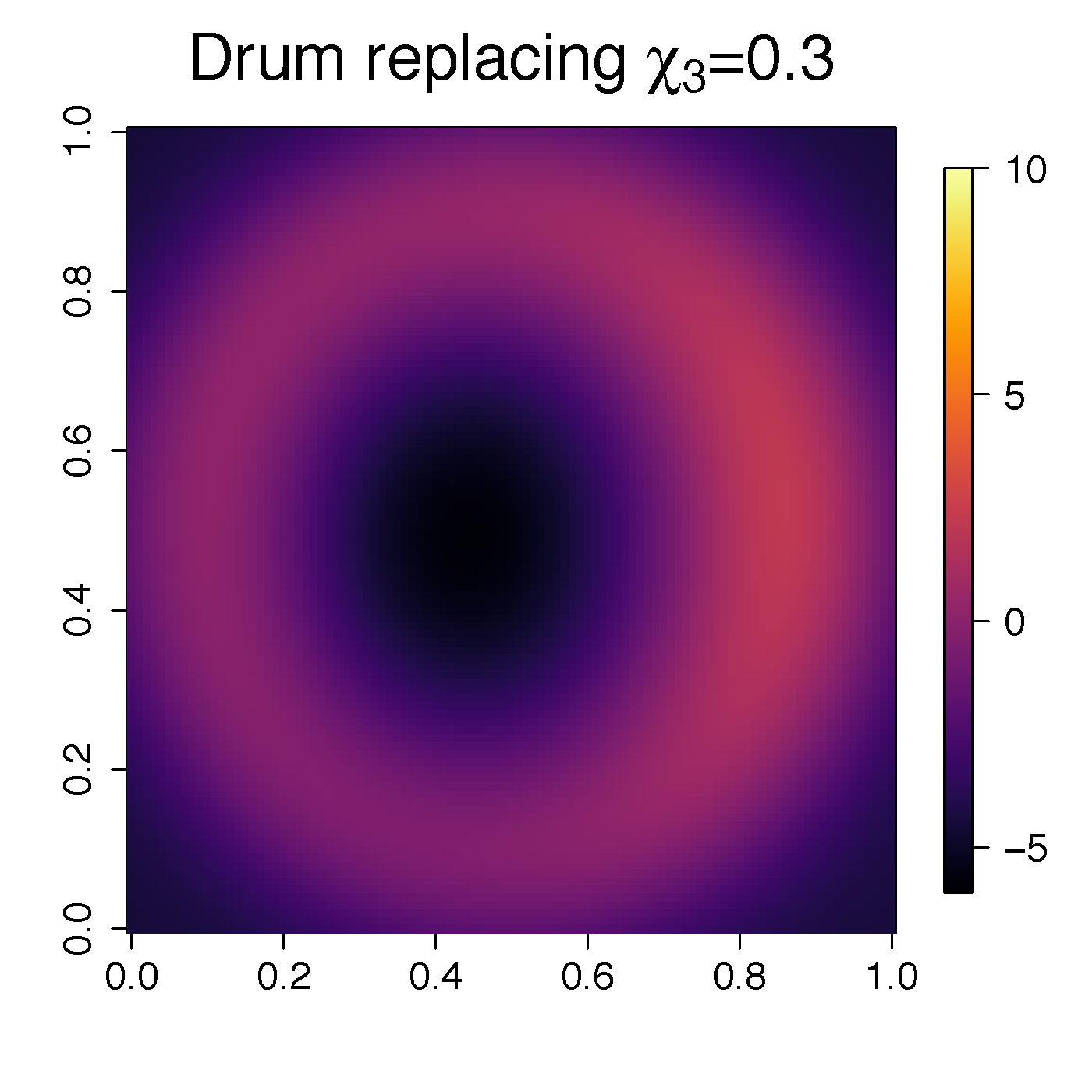} 
	\includegraphics[width=1.5in]{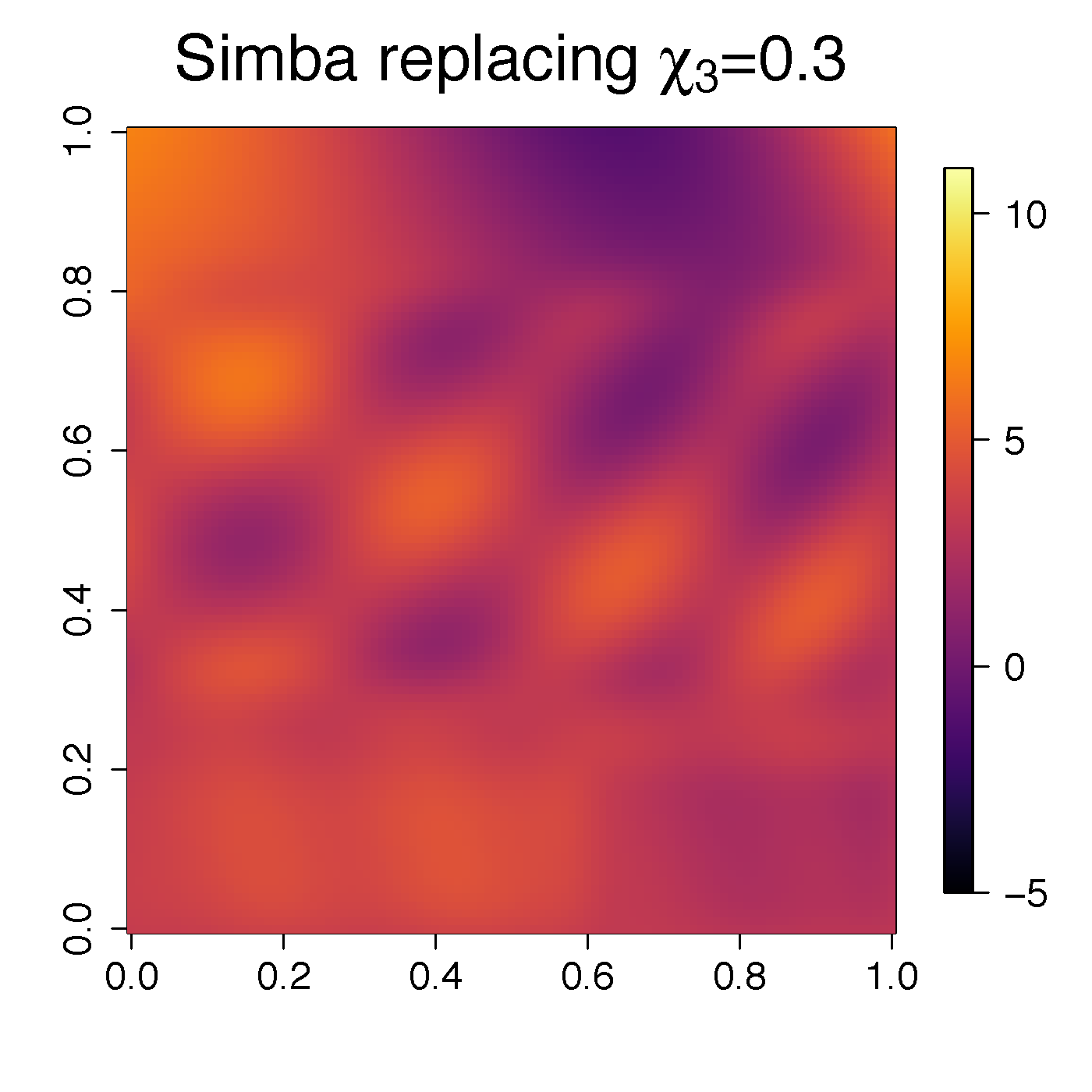} 
    \caption{ Each column shows a different 6-dimensional test function in the $x_1$, $x_2$ dimension with the other coordinates fixed.  The first row sets $x_3$ to $x_6$ to their optimal values in $\boldsymbol{\chi}$. Beach has $\boldsymbol{\chi} = 1,0.85,1,0,0,0)$; Drum has $\boldsymbol{\chi} = (0.368, .533, 0, 1, 0.555, 1)$; and Simba has $\boldsymbol{\chi} =(.523, .0999, 0, 0.298, .298, .245)$.  Plots in the second row are nearly identical to those in the first, indicating $x_6$ is locally inactive. The third row shows that $x_3$ is locally active, as changing $x_3$ led to very different behavior of $f$.}
\label{f:simba}
\end{figure}

We compared results for three different test functions in $p=15$ dimensions, named Beach, Drum, and Simba.  Although these are not conventionally high-dimensional functions, they are still sufficiently large enough to be important for practical concerns. All three functions have $6$ truly globally active variables.  The names for each function come from their visualization in the $x_1$ and $x_2$ subspace with all other variables set to $\chi_j$, their values in the global maximizer, visualized in the top row of Figure~\ref{f:simba}.  The Beach function resembles a sandy beach along a pink sea; Drum resembles an oval shaped drum; and Simba is reminiscent of the scene from Disney's ``The Lion King" \citep{simba}, where Rafiki holds Simba high up on Pride Rock, against the rolling hills and surrounding plains.  The Beach function is constructed to have a local mode in a region far away from $\boldsymbol{\chi}$. Four variables are locally active around this local mode, but only $x_1, x_2$ and $x_3$ are locally active around $\boldsymbol{\chi}$. The Drum function is primarily influenced by $x_3$, but around $\boldsymbol{\chi}$, $x_1$ through $x_5$ are all locally active. The Simba function is especially challenging to optimize, since it has a large number of local modes involving all six globally active variables. Around $\boldsymbol{\chi}$, however, only $x_1, x_2$ and $x_3$ are locally active.  

It was important to choose $n_0$ to be large enough to construct a reasonable $\hat{f}$ without being so large that  $\boldsymbol{\chi}$ would easily be known. To that end, the initial designs for Beach and Drum had $n_0 = 70$ observations, and Simba had $n_0 = 80$.  For all methods we used non-informative priors $\sigma_\mu=100$, $a_\eta=b_\eta=0.1$, $a_\theta=b_\theta=1$. We set $a_u=1$ and $b_u=10$ so that $E(u_k) = 10$ and $\text{Var}(u_k) = 100$ (see~\eqref{e:mixture}). We ran MCMC chains of length $M = 1000$, of which $m = 100$ posterior draws were used for the marginal surfaces. We set $\delta = 0.30$ and chose conservative global and local variable selection thresholds, $g = 0.05$ and $\rho = 0.02$. Of primary interest was determining the response of the true function $f$ evaluated at $\hat{\boldsymbol{\chi}}$ across $N=25$ additional runs. Because each initial designs gave a different $\xopt^0$, our performance metric was relative improvement
\begin{equation}\label{e:improve}
f(\hat{\boldsymbol{\chi}}^i) - f(\hat{\boldsymbol{\chi}}^0)\ .\
\end{equation}
To summarize across all $N = 25$ runs, we defined overall improvement $\dfrac{1}{25}\sum_{i=1}^{25} f(\hat{\boldsymbol{\chi}}^i) - f(\hat{\boldsymbol{\chi}}^0)$.

\begin{figure}[hbt]
  \centering
  \includegraphics[width= .65 \textwidth]{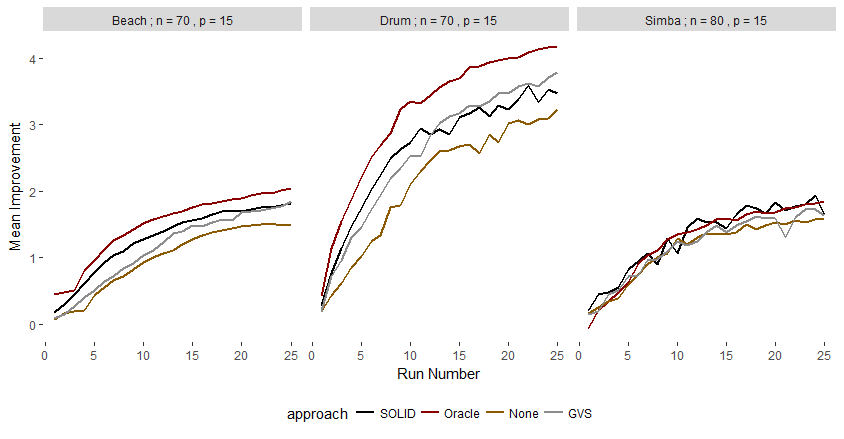} 
    \caption{ Mean relative improvement ($f(\boldsymbol{\hat\chi}^i) - f(\boldsymbol{\hat\chi}^0)$) across 100 simulations, using the three test functions for sequential runs $i \in \{1, ..., 25\}$.  Improvement at run $0$ is not shown as all approaches have zero relative improvement.}
  \label{simbaREZ}
\end{figure}

Averaging across 100 simulated initial designs, we present the mean relative improvement for each added sequential design point, for each approach and test function in Figure~\ref{simbaREZ}. As expected, Oracle performed the best on Beach and Drum, largely because it optimized across only the 6 globally active variables. Across all test functions, None performed the worst, since it always optimized over a 15-dimensional space. %Both GVS and SOLID performed global variable selection at the beginning of each run. However, SOLID optimizes $AEI$ and $\hat{f}$ over only the locally active variables, whereas GVS optimizes over the entire space of the globally active variables. 
SOLID had higher mean relative improvement than GVS for each of the first 10 runs on the Drum, and each of the first 20 runs on the Beach. For the Simba function, SOLID had higher mean improvement than GVS and Oracle for each of the first 6 runs.  In Table~\ref{r:Overall}, SOLID had significantly higher overall improvement than GVS (p-value $< 0.001$) in terms on Beach and Simba, and even outperformed Oracle on Simba ($p=0.003$).

% We present the improvement over number of runs for each setting and approach, averaged across all 100 simulated datasets, in Figure \ref{simbaREZ}. SOLID (in black) performs better than GVS on the Beach and Simba functions. In the Beach function, Oracle performs the best, achieving the highest mean improvement at every run. SOLID comes next, taking in a lead for the first 20 runs, until GVS eventually realizes similar mean improvement values by run 20. None performs the worst; even after 25 runs, the mean improvement is lower than the one obtained after 12 runs of GVS, 10 runs of SOLID, and 6 runs of Oracle.  On the Drum function, Oracle outperforms all the other approaches. Compared with GVS, SOLID takes the lead for the first 10 runs, but has mean improvement values at or below those of GVS for the latter runs. For the Simba function, SOLID was able to identify and optimize over the locally active variables, and so it outperformed all the other approaches. 

%
%The mean improvement curves are smoothest for the Beach function, primarily because of its functional form. The global maximizer is a quadratic function of $x_2$, the local maximizer is a quadratic function of $x_1$, and a third sub-optimal region has $x_1, x_2, ..., x_6$ contributing through various cosine functions involving interactions of the six variables. Instead of using discontinuous indicator functions to move from one region to another, Beach uses the CDF of a normal distribution, determined by $x_1, x_2$ and $x_3$. 

\begin{table}[htb]
\centering
\caption{The mean improvement for all approaches and test functions averaging across 100 simulated initial designs and 25 runs. A Wilcoxon rank sums test shows that SOLID performs significantly better than GVS on the Beach and Simba functions.
}
\renewcommand{\arraystretch}{1.3}
\label{r:Overall}
\resizebox{\columnwidth}{!}{%
\begin{tabular}{@{}cccccccccc@{}}
\toprule
        
& \multicolumn{2}{c}{\textbf{Oracle}}
& \multicolumn{2}{c}{\textbf{SOLID}}          
& \multicolumn{2}{c}{\textbf{GVS}}  
& \multicolumn{2}{c}{\textbf{None}}             
& 
\begin{tabular}[c]{@{}c@{}}
GVS $\ne$ SOLID
% 2-Sided Wilcoxon p-values \\ comparing GVS and SOLID
\end{tabular} \\ 
% \midrule
 \cmidrule(r){2-3}
 \cmidrule(r){4-5}
 \cmidrule(r){6-7}  
 \cmidrule(r){8-9}  
\textbf{Test Function} 
& \textbf{Mean} & {\color[HTML]{656565} Std err} 
& \textbf{Mean} & {\color[HTML]{656565} Std err} 
& \textbf{Mean} & {\color[HTML]{656565} Std err} 
& \textbf{Mean} & {\color[HTML]{656565} Std err} 
& \textbf{P-value}                                                                           \\
Beach         
& 1.49          & {\color[HTML]{656565} 0.026}  %oracle
& 1.30          & {\color[HTML]{656565} 0.023}  %solid
& 1.15          & {\color[HTML]{656565} 0.024}  %gvs
& 1.00          & {\color[HTML]{656565} 0.021}  %none
& \textless 0.001                                                                            \\
Drum          
& 3.20          & {\color[HTML]{656565} 0.037}  %or
& 2.66          & {\color[HTML]{656565} 0.038}  %solid
& 2.62          & {\color[HTML]{656565} 0.038}  %gv
& 2.14          & {\color[HTML]{656565} 0.035}  %no

& 0.122                                                                                      \\
Simba         
& 1.25          & {\color[HTML]{656565} 0.025}  %ora
& 1.39          & {\color[HTML]{656565} 0.029}  %so
& 1.19          & {\color[HTML]{656565} 0.025}  %gv
& 1.13          & {\color[HTML]{656565} 0.024}  %non

& \textless 0.001                                                                            \\ \bottomrule
\end{tabular}
}
\end{table}

\begin{figure}[htb]
	\centering
	\includegraphics[width= .65\textwidth]{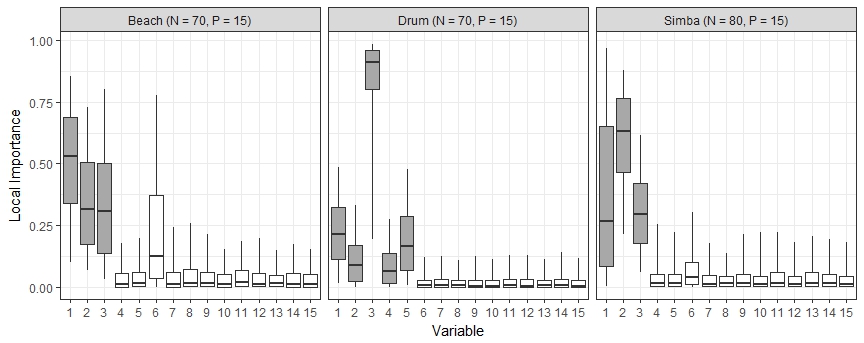}
	%   {LocalImport20Dec.jpeg}
	\caption{ $L_k$ boxplots over all 100 simulated datasets and 25 runs, where the ``whiskers" are the $5^{th}$ and $95^{th}$ percentiles. Gray boxplots correspond to the truly locally active variables.  }
	\label{localimp}
\end{figure}

SOLID was able to achieve its enhanced performance, not only by permanently removing variables through global variable selection, but by honing in on more promising lower-dimensional subspaces. Figure~\ref{localimp} shows that our proposed measure of local importance successfully captured the locally active variables. Table \ref{r:varselect} shows that across all three test functions, SOLID was optimizing over fewer variables than GVS, as defined by the number of variables explored by the $AEI$ function. %However, SOLID incorrectly includes more globally inactive variables (false positives) than GVS. As an example of how this is possible, consider the following: SOLID could optimize over 3 locally active variables and 2 globally inactive variables, whereas GVS might optimize over 6 globally active variables and 2 globally inactive variables.

 \begin{table}[!hb]
 \centering
 \renewcommand{\arraystretch}{1.3}
 \caption{The false positives are variables that are included in the design but are truly globally inactive. There are 15 total variables; at most 9 are false positives. }
 \label{r:varselect}
 \begin{tabular}{@{}lllllllll@{}}
 % \toprule
 \multicolumn{9}{l}{\textbf{Mean Number of Variables} at run 25 (across 100 simulations) }                                                                              \\ \midrule
 & \multicolumn{3}{l}{Used for Optimization }
 & \multicolumn{1}{l}{ \hphantom{A} }
 & \multicolumn{3}{l}{False Positives}  \\  
 \textbf{\begin{tabular}[l]{  @ {} l @ {} } Test Function \\ Approach \end{tabular}} 
          & Beach        & Drum         & Simba   & \hphantom{A}    & Beach       & Drum      & Simba      \\
 Oracle   & 6.00         & 6.00         & 6.00    &\hphantom{A}     & 0.00        & 0.00      & 0.00       \\
 SOLID    & 6.32         & 5.58         & 6.65    &\hphantom{A}     & 7.11        & 6.88      & 7.71       \\
 GVS      & 9.38         & 11.83        & 10.85   &\hphantom{A}     & 4.30        & 6.22      & 5.71       \\
 None     & 15.00        & 15.00        & 15.00   &\hphantom{A}     & 9.00        & 9.00      & 9.00       \\ 
 \bottomrule
 \end{tabular}
 \end{table}

% \floatsetup[table]{objectset=centering,capposition=top}

% \floatsetup[table]{objectset=centering,capposition=top}
%    \begin{table}[htb]\caption{ Mean number of variables for each approach and test function at the 25th run, averaging across 100 initial simulated datasets. False positives are the number of globally inactive classified as globally active. At most 9 of the 15 variables can be false positives. The number of optimized variables is the number of locally active variables (for SOLID) or globally active variables (for all other methods).}
%        \begin{tabular}{lllllll}
%        \toprule
%            &\multicolumn{2}{c}{\textbf{Beach}}
%            &\multicolumn{2}{c}{\textbf{Drum}}
%            &\multicolumn{2}{c}{\textbf{Simba} } 
%            \\
%            \cmidrule(r){2-3}
%            \cmidrule(r){4-5}
%            \cmidrule(r){6-7}   
%            \textbf{Approach}
%            &Optimize & False Pos
%            &Optimize & False Pos
%            &Optimize & False Pos
%            \\ 
%            \midrule
%            Oracle  & 6.00 & 0.00
%                    & 6.00 & 0.00
%                    & 6.00 & 0.00\\
%            SOLID   & 6.32 & 7.11
%                    & 5.58 & 6.88
%                    & 6.65 & 7.71 \\
%            GVS     & 9.38 & 4.30
%                    & 11.83 & 6.22
%                    & 10.85 & 5.71\\
%            None    & 15.00 & 9.00 
%                    & 15.00 & 9.00
%                    & 15.00 & 9.00\\
%        \bottomrule
%        \end{tabular}
%        \centering
%        \label{r:varselect}
%    \end{table} 
% (sd $=0.06$),(sd $=0.29$)  (sd $= 0.36$)
We also compared the methods in terms of computational costs. Oracle, which knows the set of globally active variables and does not perform any variable selection, took 1.8 hours to add 25 design points, averaged across all test functions. For every hour that Oracle took to obtain 25 new design points, None took 1.4 hours, GVS took 4.0 hours, and SOLID took 5.3 hours. Although computationally more expensive, if each evaluation of $f$ is expensive, SOLID would still be preferable to GVS and None, since it required fewer evaluations of $f$ to obtain equivalent or better estimates of $\boldsymbol{\chi}$. To see this, we compared the mean improvement value that Oracle achieves after 7 runs with the number of runs the other approaches needed to achieve at least that value (see Figure~\ref{simbaREZ}). On the Beach function, SOLID needed 10 runs, whereas GVS required 13, and None required 15. Similar patterns held for the Drum and Simba functions. %If keeping the number of function evaluations to a minimum is of greater importance than computational time, we argue that SOLID outperforms the standard methods. 

% On the Drum function: SOLID attained a higher value with 10 runs, while GVS needed 12, and None needed 16. On the Simba function, both SOLID and Oracle needed 7 runs, whereas both GVS and None needed 9.

As evidenced in Figure~\ref{simbaREZ}, the mean improvement value for GVS eventually met (on the Beach and Simba functions) or exceeded (on the Drum function) the value obtained by SOLID. The difference in how these two methods selected inputs at which to evaluate $f$ next could explain this result. By selecting inputs whose values vary in all $p$ globally active dimensions, GVS may be better able to identify truly globally inactive variables and correctly remove them from the design matrix. By removing more globally inactive variables than SOLID, GVS could eventually experience comparable or better mean improvement values. In the first few sequential runs, however, the results favored SOLID. 

\section{Analysis of Sarcos Robot Data}\label{s:robot}

The Sarcos robot dataset \citep{ vijayakumar2000} consists of $n = 44,484$ observations and $p = 21$ input variables, available at $\text{www.gaussianprocess.org/gpml/data}$. The input variables are the positions, velocities, and accelerations of seven different points on a robot arm as it draws a figure eight \citep{vijayakumar2005}. We transform the inputs such that $\mathbf{x} \in [0,1]^{21}$. The response variable $Y(\mathbf{x})$ is the first of seven joint torque measurements \citep{parker2015}. Sequential optimization requires being able to evaluate the response surface at arbitrary input values, but this is not possible with the discrete Sarcos data. Therefore, for illustration purposes, we generated data assuming the true response surface is a kernel smoothed function. For any input $\mathbf{x}$, we have 
\begin{equation}\label{surrogate}
f(\mathbf{x}) = \dfrac
{ \sum_{i \in \mathbb{S} } y(\mathbf{x}_i) K(\mathbf{x}_i, \mathbf{x})  }
{  \sum_{i \in \mathbb{S} } K(\mathbf{x}_i, \mathbf{x}) },
\end{equation}
where $\mathbb{S} = \{1, ..., n\}$, $\mathbf{x}_i$ are the observed inputs in the Sarcos dataset, and the kernel smoother is $K(\mathbf{x}, \mathbf{x}') =
\text{exp}\{- \sum_{j=1}^p h^{-2}(x_j - x'_j)^2 \}$. Based on 5-fold cross-validation minimizing the out-of-sample prediction MSE, the best bandwidth was $h=0.08272$. 
% Based on insight from \cite{parker2015}, we set $\gamma_8$ and $\gamma_{15}$ to $1081.03$ and $\gamma_j = 43.53$ for all the other inputs. These choices were based on constructing training and test sets and minimizing the prediction MSE using out-of-sample observed inputs.

With the \texttt{fields} package in \texttt{R}, we randomly selected inputs that led to space-filling designs. We included ten times as many initial design points as dimensions \citep{loeppky2009}. We considered only $15$ sequential evaluations in this analysis due to the computational demands. Using only the GVS and SOLID approaches, we evaluated the improvement (\ref{e:improve}) at each run $ i \in \{1, ... , 15\}$. We set $g = 0.15$ and $\rho = 0.01$ to provide a moderate amount of variable selection, and we set $\delta = 0.20$ to emphasize local searches. $\tau^2=0.05$. We used the same priors and number of MCMC chains and posterior samples as in Section~\ref{s:simulation}.

  \begin{figure}[htb]
  \centering
  \includegraphics[width= .65 \textwidth]{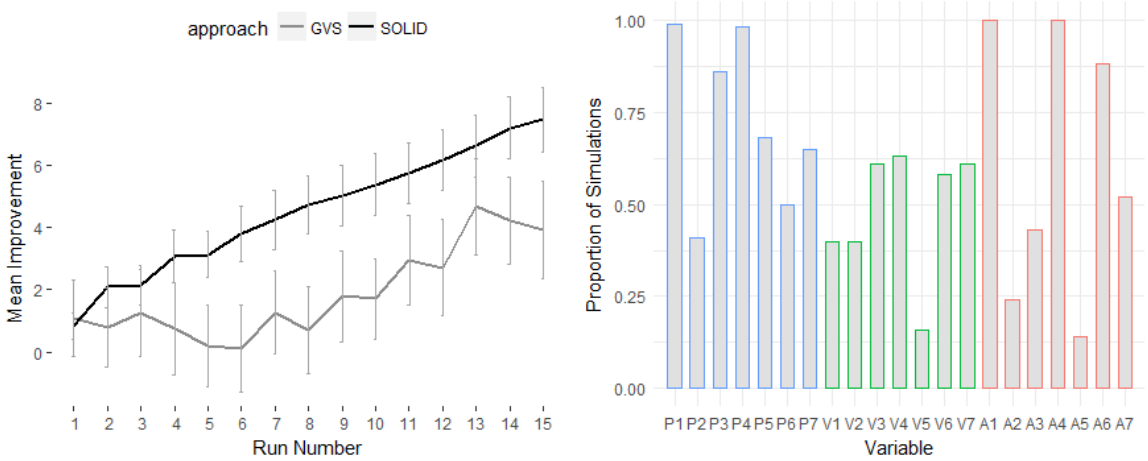}% robot_meanplot.jpeg 
    \caption{The left plot is mean relative improvement ($\pm 1$ standard error) by run across 100 simulations. The right plot is the proportion of 100 simulated datasets in which each variable was locally active at the $15^{th}$ run. The variables above correspond to the positions (P), velocities (V), and accelerations (A) of seven locations on the robot arm. % Both SOLID and GVS found all variables to globally active. SOLID considered just a few of the variables to be locally active. 
    }
  \label{robotREZ}
\end{figure}

We present results for 100 simulated datasets in Figure \ref{robotREZ}. SOLID achieved greater improvement than GVS over 100 simulated datasets at nearly every run of the sequential design. Comparing overall improvement, SOLID was significantly better than GVS (p-value $< 0.001$). %Because the candidate points for SOLID are often restricted to be in a promising area of the design space, the next inputs selected were near $\hat{\boldsymbol{\chi}}$, which in turn meant that $\xopt$ for SOLID did not change dramatically. %GVS, on the other hand, placed no restrictions on the candidate points and often allocated resources to explore the design space. Because of this greater exploration, GVS had more varied estimates of $f(\hat{\boldsymbol{\chi}})$ and took more evaluations of $f$ to obtain larger improvement values.
%The negative improvement values for the first few runs are a sign of GVS's sub-optimal re-estimations of the global maximizer.
SOLID consistently used fewer variables for optimization than GVS. We found that neither method removed any variables based on global variable selection. However, by the final run, SOLID used $15.88$ variables during its optimization and design selection, compared to $21$ for GVS. Figure \ref{robotREZ} shows the proportion of datasets with globally and locally active variables at the final run. Of the 21 input variables, SOLID identified several as locally active. At the last run, variables Position 1, Acceleration 1, and Acceleration 4 were identified as locally active in $93\%$, $100\%$, and $96\%$ of the simulated datasets. 

% There was no significant interaction effect between the $\rho$ and $\delta$ factors.

\section{Discussion}\label{s:discussion}
% it is important to be able to get the most ``bang for the buck," to borrow a phrase from the Eisenhower Administration \citep{pavelec2010}. The

When optimizing a function $f(\cdot)$ where each evaluation is expensive, one goal is to obtain the largest $f(\hat{\boldsymbol{\chi}})$ in as few evaluations of $f$ as possible. To that end, we proposed SOLID, a new method that measures local variable importance around $\xopt$ and uses this information to optimize $f$ in a sequential design. Whereas global variable selection permanently removes globally inactive variables, our local variable selection approach is flexible, adapting to the uncertainty of $\xopt$.  We tailored local variable selection to optimize the search for both the maximizer of the $AEI$ acquisition function and the global maximizer. Rather than exploring the entire $p-$dimensional space, SOLID examines only the locally active variables. In a simulation study, we found that our definition of local importance successfully captured the subset of locally active variables across three test functions.  By reducing the optimization dimension  global and local variable selection, our SOLID algorithm was able to achieve higher $f(\hat{\boldsymbol{\chi}})$ values compared to the standard methods, for a fixed number of sequential evaluations. %Reducing the dimension of the optimization using global and/or local variable selection led to better performance.%, especially when compared to the approach that performed no variable selection at all. 

%Because optimizing the $AEI$ surface is difficult, SOLID may achieve greater performance than the other methods due to restricting the range of the search space.

There are several ways that SOLID could be further improved. One reviewer pointed out that the presence of locally active variables could lead to non-stationary behavior in the response surface.  Depending on the nature of non-stationarity, this could result in the estimated GP spatial range parameters, which assume a stationary covariance, as poor indicators of a variable's global importance.  It is our intention that any variable that influences $f$ anywhere in the input space be classified as globally active.  Using a GP model with a non-stationary covariance function is the next step, though computational costs would increase and our definition of globally active would need to be modified. On the other hand, a non-stationary covariance function may not be necessary. The three test functions in Section~\ref{s:simulation} exhibit non-stationary behavior, yet the SOLID behaves well and rarely drops a globally active variable.  It is also possible that a stationary covariance function would still able to predict $f$ in a subregion near $\boldsymbol{\chi}$, which is one reason why we chose our local selection criterion to be based on prediction comparisons instead of directly inspecting the estimates of the spatial range parameters.

% As evidenced by the toy example, given a fixed number of runs, the current implementation of SOLID can lead to better estimates of $\boldsymbol{\chi}$ as more evaluations of $f(\cdot)$ are made, and yet still be unable to describe the entire response surface well. Investigating which choices of covariance functions could improve SOLID's performance is an area for future research. 

% Furthermore, the success of any optimizing algorithm hinges on the initial design. To prevent missing locally active variables, we used space-filling designs, which offer the best chance (but no guarantee) that the initial design will enable the surrogate function to identify the locally active variables and \textit{not} mistakenly consider them globally inactive. It is also important to choose conservative variable selection thresholds to avoid permanently removing variables that actually matter in optimizing the response surface. 

One limitation of SOLID is its required computations to estimate local importance and its utilization of MCMC to estimate $f$. In instances where the underlying function is inexpensive to evaluate, it would be faster to use conventional sequential design approaches. Additionally, for experiments involving, say, more than 50 variables, the MCMC algorithm presented here for global and local variable selection could become excessively slow and other methods would be preferable, such as the random embedding approach \citep{wang2016} or by specifying an additive model \citep{kandasamy2015}. That said, SOLID's local variable selection could be used for many functions $f$, without needing to know which or how many variables are locally active.   An area of future work would be to incorporate aspects of these other methods within the SOLID framework, perhaps to perform fast initial screening.%SOLID's performance on the robot dataset (Section \ref{s:robot}) suggests that this new method is valuable, even for functions that were not designed \textit{a priori} to have locally important variables. %By incorporating local variable importance to reduce the dimension and range of the search space, improved performance is possible, especially early on.

% \section*{Acknowledgements}
% This work was supported by National Science Foundation grants DGE-1633587 and DMR-1535082. 

\bibliographystyle{asa}
\bibliography{des}

\section{Appendix}
\subsection{MCMC Details}\label{s:mcmc}
We use Metropolis-Hastings within Gibbs sampling to obtain posterior samples of $\bTheta$. For convenience, we reparameterize to the total precision (inverse variance), $\eta = (\sigma^{2} + \tau^{2})^{-1}$, and proportion of variance from the response surface, $r =  \sigma^2\eta$. Using the parameterization in Section~\ref{s:GVS}, let 
\begin{equation}\label{vw}
V( \bx, \bx') =
\dfrac{1}{\eta} \left[
r K(\bx, \bx') + (1-r) 1_{\{ \bx = \bx'\}} \right]
\equiv \dfrac{1}{\eta} W(\bx,\bx').
\end{equation}
Denote $\frac{1}{\eta}\mathbf{W_X}$ as the $n$ x $n$ covariance matrix corresponding to $\mathbf{X}$. The log likelihood is
\begin{equation}
\log L \left(\by  \mid \bTheta, \bX \right) 
= -\dfrac{n}{2}\ln(2\pi) - \dfrac{1}{2}\ln |\dfrac{1}{\eta} \mathbf{W_X}|
-\dfrac{\eta }{2}(\by - \mu \mathbf{1}_n )^T \bW_X^{-1} (\by-\mu \mathbf{1}_n ).
\end{equation}

The full conditional distributions of $\mu$, $\eta$, $\theta$, and $b_k$ are conjugate, and so these parameters are updated by sampling from their full conditional distributions
\begin{eqnarray}
\eta \mid \text{rest} &\sim& \text{Gamma}\left( 
\dfrac{n}{2} + a_\eta \ \ , \ \  
b_\eta + \dfrac{1}{2}\left[( \mathbf{y} - \mu \mathbf{1}_n)^T \mathbf{\bW_X}^{-1} ( \mathbf{y} - \mu \mathbf{1}_n )\right] \right)\\
\mu \mid \text{rest} &\sim&
\mbox{Normal}\left( 
\frac{\eta\mathbf{1}_n^T\bW_X^{-1}\by}{\sigma_\mu^{-2}+\eta w}
\ \ , \ \ 
\dfrac{1}{\sigma_\mu^{-2}+\eta w}\right)\nonumber\\
\theta \mid \mbox{rest} &\sim& \text{Beta}\left(\alpha_\theta +  \sum_{k=1}^p b_k \ , \ b_\theta + p -  \sum_{k=1}^p b_k \right)\nonumber\\
b_{k} \mid \mbox{rest}&\sim& \text{Bernoulli}\left( \dfrac{p_{k1} }{ p_{k1} + p_{k0}}\right)\nonumber\ ,\
\end{eqnarray}
where $w = \mathbf{1}_n^T\bW_X^{-1}\mathbf{1}_n$ and $p_{k\ell} \equiv p( \by \mid b_k = \ell,\bTheta_{(-k)}) p(b_k = \ell \mid \theta)$ for $\bTheta_{(-k)} = \bTheta_k/\{b_k\}$. 

We implement the Metropolis-Hastings algorithm \citep{hastings1970} to update $r$ and $u_1,...,u_p$. The variance ratio $r$ is sampled using the Metropolis-Hastings algorithm with a $\text{Beta}(10,1)$ proposal distribution. For each $k \in \{1, ...,p\}$, if $b_k = 0$ then $u_k$ is updated from its prior, otherwise it is updated using a Metropolis-Hastings step with a sliding uniform candidate distribution, conditioned on the current value of $u_k$, and designed to propose smaller values of $u_k$ corresponding to smoother surfaces. Specifically $$u_k \sim \text{Uniform}
\left( 
\max\{0, u_k - 50\epsilon(u_k) \}
\ , \  
u_k + \epsilon(u_k)
\right),$$ where

\begin{equation}\label{propdist}
\epsilon(u_k) \equiv
\left\{
\begin{array}{ll}
\min\{50,  u_k h\} & u_k\ge 30 \\
\max\{1, u_k  h\} & 0 \le u_k < 30 
\end{array} 
\right.
\end{equation}
and $h \sim\text{Unif}(1/2,2)$.
This proposal distribution is used because it ensures candidates are positive and its candidate distribution's variance increases with the current value of $u_k$.

At each sequential step $i \in \{0, 2, ... , N\}$, we estimate $\hat{\boldsymbol{\chi}}^i$ using all $p$ globally active variables and the quasi-Newton optimizer, L-BFGS-B \citep{byrd1995}, ensuring the maximizer is contained in the design space $[0,1]^p$. We input the marginal predicted function, $\hat{f}$ and its marginal gradient %(Equation~\ref{e:marginal})
along with the previous $\xopt$ (if available) and the four design points with the largest observed responses. This multi-start search is to prevent suboptimal convergence to a local optimal.  

After obtaining $\xopt$, local variable importance is assessed using the posterior draws $\xopt_t$, and the next design point $\mathbf{x}^*$ is chosen from the $\mathcal{R}^\mathbf{A}$ or $\mathcal{R}^\delta$ restricted spaces.  We again used line searches to optimize $AEI$ within these two restricted spaces but other optimization algorithms are possible, including the genetic algorithm and derivative optimizer \texttt{genoud} \citep{mebane2011}. 
% , and it is particularly important for smaller initial designs, since the estimated global maximizer will tend to be near an observed design point. 
%each locally active variable $k \in\mathbf{A}$ explores $\mathcal{R}_k = [0,1]$, and 
In Step 9 of Algorithm 2, we restrict the search space of $\xopt$ to lie in $\mathcal{R}^\mathbf{A}$, where each locally inactive variable $j \in \mathbf{A}^c$ is fixed at the corresponding entries of $\xopt$. This leads to a refined estimate of $\xopt$, at the end of the each sequential run. 

\subsection{SOLID Local Tuning Parameters}
Two tuning parameters influence the decision to declare a variable locally active or inactive. The $\delta$ parameter in (\ref{predpoints}) affects the spread of the prediction points about $\hat{{\boldsymbol{\chi}}_t}$, and so the choice of $\delta$ affects $L_k$.  Smaller values of $\rho$ allow for a larger number of variables to be declared locally active. It is possible to choose $\rho$ after each run such that a fixed proportion of the globally active variables are considered locally active. One could reason that in ``some region of the design space, only a small number of factors are actively influencing the response"  \citep{myers2016}.

\begin{table}[htb]
\centering
\caption{A sensitivity analysis on 100 simulated datasets, showing mean improvement after 7 runs, for 6 combinations of $\rho$ and $\delta$. All initial designs use $21$ dimensions and $10 \times 21 = 210$ design points.  }
\renewcommand{\arraystretch}{1.3}
\begin{tabular}{@{}ccc@{}}
\toprule
\multicolumn{3}{c}{Mean Improvement (Standard Error) } \\ \midrule
\multicolumn{1}{r}{ Settings              }   & {  $\delta =$ 0.2 }  & {  $\delta =$ 0.6 }    \\
$\rho = $0.01    & 6.55 (1.21) & 9.17 (1.50) \\
$\rho = $0.05    & 6.32 (1.32) & 8.63 (1.42) \\
$\rho = $0.15    & 3.60 (1.38) & 5.56 (1.05) \\
\bottomrule
\end{tabular}

\label{robotSEN}
\end{table}

To examine how sensitive SOLID's performance was to different specifications of the $\rho$ local variable selection threshold and the $\delta$ radius for local importance, we considered 100 simulations using a two-factor crossed design. Because so few variables were declared to be globally inactive, we disabled the global variable selection feature in the sensitivity analysis. We set the initial design to have size $n = 210$ and considered the improvement after 7 runs, limiting the number of runs due to computational costs. Results in Table \ref{robotSEN} show that a larger radius ($\delta = 0.60$) and conservative threshold ($\rho = 0.01$ or $\rho = 0.05$) provided for the best performance. As long as a conservative local variable selection threshold was chosen, the results were not too sensitive to $\rho$. 

\end{document}

% Supplementary Materials Follow
\begin{center}
	\LARGE \textbf{Supplementary Materials}
\end{center}
\section{MCMC details}
We use Metropolis-Hastings within Gibbs sampling to obtain posterior samples of $\bTheta$. Using the parameterization in Section~2 of the main document, let 
\begin{equation}\label{vw}
V( \bx, \bx') =
\dfrac{1}{\eta} \left[
r K(\bx, \bx') + (1-r) 1_{\{ \bx = \bx'\}} \right]
\equiv \dfrac{1}{\eta} W(\bx,\bx').
\end{equation}
Denote $\frac{1}{\eta}\mathbf{W_X}$ as the $n$ x $n$ covariance matrix corresponding to $\mathbf{X}$. The log likelihood is
\begin{equation}
\log L \left(\by  \mid \bTheta, \bX \right) 
= -\dfrac{n}{2}\ln(2\pi) - \dfrac{1}{2}\ln |\dfrac{1}{\eta} \mathbf{W_X}|
-\dfrac{\eta }{2}(\by - \mu \mathbf{1}_n )^T \bW_X^{-1} (\by-\mu \mathbf{1}_n ).
\end{equation}

The full conditional distributions of $\mu$, $\eta$, $\theta$, and $b_k$ are conjugate, and so these parameters are updated by sampling from their full conditional distributions
\begin{eqnarray}
\eta \mid \text{rest} &\sim& \text{Gamma}\left( 
\dfrac{n}{2} + a_\eta \ \ , \ \  
b_\eta + \dfrac{1}{2}\left[( \mathbf{y} - \mu \mathbf{1}_n)^T \mathbf{\bW_X}^{-1} ( \mathbf{y} - \mu \mathbf{1}_n )\right] \right)\\
\mu \mid \text{rest} &\sim&
\mbox{Normal}\left( 
\frac{\eta\mathbf{1}_n^T\bW_X^{-1}\by}{\sigma_\mu^{-2}+\eta w}
\ \ , \ \ 
\dfrac{1}{\sigma_\mu^{-2}+\eta w}\right)\nonumber\\
\theta \mid \mbox{rest} &\sim& \text{Beta}\left(\alpha_\theta +  \sum_{k=1}^p b_k \ , \ b_\theta + p -  \sum_{k=1}^p b_k \right)\nonumber\\
b_{k} \mid \mbox{rest}&\sim& \text{Bernoulli}\left( \dfrac{p_{k1} }{ p_{k1} + p_{k0}}\right)\nonumber
\end{eqnarray}
where $w = \mathbf{1}_n^T\bW_X^{-1}\mathbf{1}_n$ and $p_{k\ell} \equiv p( \by \mid b_k = \ell,\bTheta_{(-k)}) p(b_k = \ell \mid \theta)$ for $\bTheta_{(-k)} = \bTheta_k/\{b_k\}$. 

We implement the Metropolis-Hastings algorithm \citep{hastings1970} to update $r$ and $u_1,...,u_p$. The variance ratio $r$ is sampled using the Metropolis-Hastings algorithm with a $\text{Beta}(10,1)$ proposal distribution. For each $k \in \{1, ...,p\}$, if $b_k = 0$ then $u_k$ is updated from its prior, otherwise it is updated using a Metropolis-Hastings step with a sliding uniform candidate distribution, conditioned on the current value of $u_k$, and designed to propose smaller values of $u_k$ corresponding to smoother surfaces. Specifically $$u_k \sim \text{Uniform}
\left( 
\max\{0, u_k - 50\epsilon(u_k) \}
\ , \  
u_k + \epsilon(u_k)
\right),$$ where

\begin{equation}\label{propdist}
\epsilon(u_k) \equiv
\left\{
\begin{array}{ll}
\min\{50,  u_k h\} & u_k\ge 30 \\
\max\{1, u_k  h\} & 0 \le u_k < 30 
\end{array} 
\right.
\end{equation}
and $h \sim\text{Unif}(1/2,2)$.
This proposal distribution is used because it ensures candidates are positive and its candidate distribution's variance increases with the current value of $u_k$.

% Since the values of each $u_1,..., u_p$ can be of different orders of magnitude, by automatically adjusting the scale of proposed $u_k$ values, we do not need to manually specify the width of the proposal distribution. An example appears in Table \ref{proposal}.

% \begin{table}[!ht]
% \centering
% \caption{Examples of Proposal Values}
% \label{proposal}
% \begin{tabular}{@{}cccc@{}}
% \toprule
% {\color[HTML]{333333} \begin{tabular}[c]{@{}c@{}}Current Value \\ $u$\end{tabular}} & {\color[HTML]{333333} \begin{tabular}[c]{@{}c@{}}Random Uniform\\  $h \sim$ U(.5, 2)\end{tabular}} & {\color[HTML]{333333} \begin{tabular}[c]{@{}c@{}}Margin Term\\ $\epsilon_u$\end{tabular}} & {\color[HTML]{333333} Range of Proposal Values} \\ \midrule
% 0 & 1.00 & 0.95 & [0, 0.95] \\
% 0.08 & 0.75 & 0.95 & [0, 0.97] \\
% 0.35 & 2.00 & 0.7 & [0, 1.05] \\
% 5 & 1.97 & 9.85 & [0, 14.85] \\
% 15 & 0.55 & 8.25 & [6.75, 23.25] \\
% 50 & 1.30 & 65 & [0, 115] \\
% 150 & 1.00 & 100 & [50, 250] \\
% 300 & 0.50 & 100 & [200, 400] \\
% 800 & 1.60 & 100 & [700, 900] \\
% 1500 & 1.04 & 100 & [1400, 1600] \\ \bottomrule
% \end{tabular}
% \end{table}

\newpage

\section{Details on SOLID Algorithm}

\subsection{Maximizing AEI Using Local Information}\label{s:maxAEI_details} 

Let $\mathbb{A}$ denote the estimated set of locally active variables at $\hat{{\boldsymbol{\chi}}}$ for the current design.  Likewise, let $\mathbb{A}_{\hat{\boldsymbol\chi}}$ denote the active subspace of $[0,1]^p$ where inactive variables are projected to their optimal inputs in ${\hat{\boldsymbol\chi}}$.  Correctly specifying $\mathbb{A}_{\hat{\boldsymbol\chi}}$ can greatly reduce the difficulty of optimizing the $AEI$ surface by focusing on a subspace exhibiting substantial variation of the response surface near the estimated global maximizer.  We next detail a strategy for maximizing AEI using this local importance information.

It is nontrivial to find the optimal input for $AEI$ since the $AEI$ surface could be multi-modal.  We simplify the optimization by reducing the dimension to $\mathbb{A}_{\hat{\boldsymbol\chi}}$, fixing all inactive variables to their values estimated by $\hat{\boldsymbol\chi}$.  Even still, this space may still be large, and there is the question of whether we should explore the entire subspace or just a neighborhood of the subspace centered around $\hat{\boldsymbol\chi}$.  After all, if $\hat{\boldsymbol{\chi}}$ is truly near the global maximizer, restricting the search for the $AEI$ maximizer to a neighborhood of $\hat{\boldsymbol{\chi}}$ could be advantageous.

First, we evaluate $AEI$ over $C$ candidate points that form a maximin LHS design \citep{kleijnen2015} in $\mathbb{A}_{\hat{\boldsymbol\chi}}$.  Denote this candidate set by $\mathbf{C}_\mathbb{A}$.  We then construct another set of $C$ candidate points, denoted $\mathbf{C}_{\mathbb{A} \delta}$, that form a maximin LHS design in a $\delta$-neighborhood of $\hat{{\boldsymbol{\chi}}}$ contained in $\mathbb{A}_{\hat{\boldsymbol\chi}}$.
% Furthermore, since we have posterior draws of the optimal input in each dimension, we can choose to shrink the area of exploration. 
To create this neighborhood, for each $k \in \mathbb{A}$, we set the restricted lower and upper limits to be
$\max{\{ 0 , \min{  ( \hat{\chi}_{k1}, ...,  \hat{\chi}_{km} )}  - \delta\}}$ and 
$\min{\{1 , \max{  (  \hat{\chi}_{k1}, ...,  \hat{\chi}_{km} )}  + \delta\}}$ using the $m$ posterior draws of each estimate of the $k$-th coordinate of the global maximizer, denoted $\hat{\chi}_{kt}$ for the $t$-th draw. SOLID evaluates $AEI$ at the $2C$ candidate points in $\mathbf{C}_{\mathbb{A}\delta}$ and $\mathbf{C}_\mathbb{A}$. Guidance for choosing $\delta$ is given in Section \ref{s:tuning}.

At this stage, it is possible to combine both sets of candidate points into one large set, $\mathbf{C}$. In practice, we find that the candidate points with the greatest $AEI$ are often all in one of the two sets. In our implementation, whichever set has the largest $AEI$ becomes the set of candidate points $\mathbf{C}$ for the next step. Conceptually, this helps us see if SOLID is honing-in on a restricted space $\mathbf{C}_{\mathbb{A}\delta}$ or exploring the larger space $\mathbf{C}_{\mathbb{A}}$. %This way, each iteration of the sequential design uses the same number $C$ of candidate design points.
% Using these ${\hat{\boldsymbol{\chi}}}$ draws instead of the optimal input ${\bar{\boldsymbol{\chi}}}$ reduces the chances of being confined to too small a search region.

%\subsection{ Selecting the Next Design Point}\label{s:nxtpnt}
%Once a set of candidate points is chosen, 
Next, we choose the five candidate points with the greatest $AEI$ and perform local line searches in the direction of the $|\mathbb{A}|$-dimensional $AEI$ gradient using only the locally active dimensions. The $AEI$ gradients of the locally inactive variables are set to zero. We place two restrictions on the local line search. First, we require that the line search lie within a $p$-dimensional ball of radius $\delta$ centered at $\mathbf{x^*}$, the starting point. Second, whenever the line search proposes a coordinate outside the design space, that coordinate is set to equal the boundary value and the corresponding coordinate of the gradient is set to zero. %In two dimensions, if the horizontal gradient hits the boundary, the subsequent design points of that line search would either slide up or down the boundary, depending on whether the vertical gradient was positive or negative. 
This approach was inspired by the more rigorous approach of \cite{rosen1960}. Once the five line searches are complete, the one with the largest $AEI$ is chosen as the next design point.

\subsection{Choosing the $\delta$ Tuning Parameter}\label{s:tuning}

The $\delta$ term handles several aspects of SOLID's performance. First, it controls the proportion of the design space that is used to establish local importance. Second, $\delta$ determines how far away the restricted candidate points $\mathbf{C}_{\mathbb{A}\delta}$ can be from the estimated maximizer $\hat{\boldsymbol{\chi}}$. Third, it limits how far the local line searches for $AEI$ can go. Simulation studies show that local importance is fairly constant for $\delta>0.50$. In choosing a $\delta$ value, an important requirement is that the mean of the prediction points (whose distance from the global maximizer depends on $\delta$, decreases as $\delta$ increases. As an alternative to specifying $\delta$ directly, one could choose $\delta$ such that the mean of the prediction points is, say, $80\%$ of the estimated global maximum. 

\newpage

\begin{algorithm}[H]
	\caption{SOLID: Sequential Optimization in Locally Important Dimensions}\label{SOLIDalg}
	\begin{algorithmic}[1]
		\Procedure{ Sequential Design}{}
		\State {Create an initial maximin LHS$(n,p)$ design, $\mathbf{X}$}
		\State Evaluate $Y(\mathbf{X}) = f(\mathbf{X}) +\epsilon$
		\State {Fit a Gaussian process and obtain $\boldsymbol{\Theta}_1, ..., \boldsymbol{\Theta}_M$ draws from the joint posterior distribution}
		\For{step $i\in\{1,...,N\}$}
		\State {Perform global variable selection, \Call{GVS}{$\boldsymbol{\Theta}, \kappa$}}
		\State Update parameter estimates $\boldsymbol{\Theta}_1,...,\boldsymbol{\Theta}_M$ after global variable selection
		\State Estimate set of locally active variables $\mathbb{A}$ using \Call{LVS}{$ \boldsymbol{\Theta}, i$}
		\State Construct a matrix of candidate points $\mathbf{C}$ using \Call{Candidates}{$\mathbb{A}, \boldsymbol{\Theta}, \delta$}
		\State Choose the design point $\mathbf{x^{**}}$ that maximizes $AEI$ using \Call{AEI}{$
			% \mathbb{A},
			\mathbf{C},\boldsymbol{\bar{\Theta}}$}
		\State Augment $\mathbf{x}^{*}$  to $\mathbf{X}$ and $Y(\mathbf{x}^{*})$ to $Y$
		\State Update parameter estimates again
		\State Estimate the optimal input $\hat{\boldsymbol{\chi}}$. 
		\EndFor
		\EndProcedure
		\Statex
		
		\Function{GVS}{$ \boldsymbol{\Theta}, \kappa $}
		% \State Specify cost of ``false-active" ($\lambda_c$) and ``false-inactive" ($\lambda_h$) 
		\For{variable $k \in \{1,...,p\}$} 
		% \State Estimate $ b_k \equiv using $$
		\State{Drop $X_k$ from the design matrix  $\iff Pr( \gamma_k > 0 \mid \mathbf{y}, \boldsymbol{\bar{\Theta}})  < \kappa $}
		\EndFor
		\EndFunction
		\Statex
		\Function{LVS}{$ \boldsymbol{\Theta}$, i}
		\State Randomly sample $m < M$ posterior draws of the optimal design point $\hat{\boldsymbol{\chi}}_1, ... ,  \hat{\boldsymbol{\chi}}_m$
		\For{ $t \in \{1,...,m\}$}
		\State Make predictions $\hat{f} \mid \boldsymbol{\Theta}_t$ at $q$ points using $\mathbf{X}$ 
		\For{ variable $k \in \{1,..,p\}$}
		\State Set $\gamma_k = 0$ and make alternative predictions, $\hat{f}^{k}$ at the same $q$ points
		\State Calculate the local importance, $L_{t k} \equiv 1 - \text{Corr}(\hat{f}, \hat{f}^{k} \mid \boldsymbol{\Theta}_t)^2$
		\EndFor
		\EndFor
		\State Summarize across posterior draws and calculate $L_k = \text{mean}(L_{1k},...,L_{mk})$ for $k \in \{1,...,p\}$. 
		\State Let $\mathbb{A}_i = \{ k: L_k \ge \rho \mid \rho\in(0,1) \}$ be the set of locally active variables at step $i$.
		% \State If no variables meet this criteria, set $\rho = \max\{L_1, ..., L_p\}$ such that at least one variable is considered locally active.
		\Return{ $\mathbb{A}_i$}
		\EndFunction
		\Statex
		\Function{Candidates}{$\mathbb{A},\boldsymbol{\Theta}, \delta$}
		\State Create $\mathbf{C}_{\mathbb{A}\delta}$ and fill the locally active dimensions with a maximin LHS($C$, $|\mathbb{A}|$) design, rescaled such that the bounds are restricted to be between $\max{\{ 0 , \min{  ( \hat{\chi}_{k1}, ...,  \hat{\chi}_{km} )}  - \delta\}}$ and 
		$\min{\{1 , \max{  (  \hat{\chi}_{k1}, ...,  \hat{\chi}_{km} )}  + \delta\}}$ for $k \in \mathbb{A}$, using the estimated optimal inputs $\hat{\boldsymbol{\chi}}_\ell$ for $\ell \in \{1,...,m\}$
		\State Fix the remaining dimensions at the estimated optimal input $\hat{\chi}_k$ for $k \not\in\mathbb{A}$
		\State Create $\mathbf{C}_\mathbb{A}$, a maximin LHS($C, p)$ design $\in [0,1]^p$, and calculate the largest $AEI$ on $\mathbf{C}_\mathbb{A}$. 
		\If {  
			$\max_{\mathbf{x} \in \mathbf{C}_\mathbb{A} } AEI(\mathbf{x}) > \max_{\mathbf{x} \in \mathbf{C}_{\mathbb{A}\delta}} AEI(\mathbf{x})$ 
		}{ 
			$\mathbf{C} \equiv \mathbf{C}_\mathbb{A}$ } 
		\Else {
			$\mathbf{C} \equiv \mathbf{C}_{\mathbb{A}\delta}$
		}
		\Return $\mathbf{C}$
		\EndIf
		\EndFunction
		\Statex
		\Function{AEI}{$\mathbb{A},\mathbf{C},\boldsymbol{\bar{\Theta}}$}
		\State Evaluate AEI at each of the candidate points $\mathbf{C}$
		\State Set $\mathbf{x}^{(i)} = \arg \max_{\mathbf{x} \in \mathbf{C}} AEI(\mathbf{x})$
		\If{ $k \in \mathbb{A}$  }{ compute $g_k$, the $k^{th}$ component of the gradient of AEI($\mathbf{x^*}$)}  \Else{ $g_k =0$}
		\EndIf
		\State Choose the 5 design points with the largest $AEI$
		\State For each of the 5 design points (WLOG: $\mathbf{x}^{(i)}$), do five line searches spanned by $\mathbf{x}^{(i)} + t \mathbf{g}$ for different step multipliers, $t \in \left[0, {\delta}{(g_1^2 + ... + g_p^2)}^{-1/2}\right]$, where the bounds keep the line searches within a radius $\delta$ of the starting point, and each subsequent line search beings where the previous one ends.
		\Return {$\mathbf{x}^{**} \equiv \arg \max_{t} AEI(\mathbf{x}^{(i)} + t \mathbf{g})$  }
		\EndFunction
	\end{algorithmic}
\end{algorithm}

\newpage

\section{R Code for Simulation Test Functions}
\begin{verbatim}
truth <- function( X , P, name = 'simba'){
  # P should be the number of globally active variables
  # for the test function to work.
  # This section of the code keeps track of which
  # variables are actually used. If X3 is required, but 
  # X does not have a column called "X3," then the "truth"
  # function will input a random number for that variable.
  if( !is.null( colnames(X) )){
    
    neednames = paste("X",1:P, sep  ="")
    xnames = colnames(X)
    
    ## unless variables given, we use random uniform entries
    XX = matrix( runif(length(X)*P), nrow(X), P)
    
    XX[, is.element(neednames, xnames)  ] = X
    X <- XX
  } else # if we know the column names, use them
    
    if( is.null(colnames(X))  & ncol(X) < P ){
      
      need = P - ncol(X)
      XX = matrix( runif(length(X)*P), nrow(X), P)
      
      colnames(XX) = paste("X", 1:P , sep="")
      XX[, 1:ncol(X)] = X
      X <- XX
    }# otherwise, we just require dimensions to match
  if(name == 'toy'){
    
    x1 = X[,1] ; x2 = X[,2];
    
    Y = 10*(x2)^2*pnorm( 10*(x1 - .4 ) ) + 
      sin(5*pi*( x2 - x1^2 ) - x1*x2 )*pnorm( 10*(.4 - x1 ) )
    
    return(Y)
    
  } else # end toy
  
  if(name == 'drum'){
    
    
    x1 = X[,1] ; x2 = X[,2]; x3 = X[,3];
    x4 = X[,4] ; x5 = X[,5]; x6 = X[,6]
    
    inner = (6 - x3/4)*cos(  4*pi* (x3-.5)   )*(
      dnorm(   11*( (x1 -.5)^2 + (x2-.5)^2 + .25  )^2 )
    )     *(1 - 3*(x1-.3)^2)
    
    middl =   (1 + 2*x4)*sin(  2*pi* x4* (x5-.3)  )*(
      pnorm(     6*( (x1 -.5)^2 + (x2-.5)^2 - 0.13 ))
      *pnorm(   -8*( (x1 -.5)^2 + (x2-.5)^2 - 0.11 ))
    )*     (   1 + x4^2 + x5^2*(x2-.2)  )
    
    outer =    (1 - 2*x5)*cos(  2*pi* (x5) * (x4+.5)   )*(
      pnorm(     8*( (x1 -.5)^2 + (x2-.5)^2 - 0.2 )   )
    )*(       1 - x6^2 - x5^2*(x1+.2)   )
    
    
    Y = (inner + middl + outer) *(10/2.032078) - 4.4831
    
    
    return(Y)
    
  } else # end drum
  
  
  if(name == 'beach'){
    
    x1 = X[,1] ; x2 = X[,2]; x3 = X[,3];
    x4 = X[,4] ; x5 = X[,5]; x6 = X[,6]
    
    sdz = .2 + (2 + x6 + x5 - 1.5*x1)*(3 + x4 - x3 - x2)/12
    
    bumps = ( 
      
      5*sin(6*pi*x1*x6)*(x3^2 +1)-( x2^2 + 4 - x1*x2/(x3 - 7) + x4*(x5 - .3))^2*  
        cos( 4*pi*x1*x3^2 )^10*(x1*x2^2 - .5)*(x2*x6 - .5)*(x5 - .5)
      
    )*
      (  pnorm( 10*( .8 - x3 ), sd = sdz )
         *pnorm( 1*(  x1 - .1 ), sd = sdz )
         *pnorm( 1*( x2 - .1 ), sd = 2*sdz )
         # *pnorm( 1*( x3 - .1 ), sd = 2*sdz )
      )
    
    horiz = ( 10.5 - 30*(x1 - .3)^2  )*
      (  pnorm( 1*( .2 - x3  ), sd = sdz )
         *pnorm( 10*( .3 - x2 ), sd = sdz ) )
    
    vert = ( 10.5 - 30*(x2 - .85)^2  )* 
      (    pnorm( 5*( x3 - .8 ), sd = sdz )
           *pnorm( 10*( x1 - .8 ), sd = sdz ) )
    
    Y = bumps + horiz + vert -.97013 + .470418
    
    
    return(Y)
    
  } else # end beach
  
  if(name == 'simba'){
    
    x1 = X[,1] ; x2 = X[,2]; x3 = X[,3];
    x4 = X[,4] ; x5 = X[,5]; x6 = X[,6]
    
    Y = 3.14749 + sin(  2*pi*(x1^2 - 2*x2*(1 + x3) ) )*
      ( pnorm( 30 *(x2 -.3))  + pnorm(30*(.8 - x2)) - 1 )*
      2*sin( 4*pi*x1 + 3*pi*(1+x3) + 2*pi*(x4 + x5) + 3*pi*(1 + x6))  +
      
      (  4 + 6*(x1) )*
      ((  pnorm( 30*(x2 - .0)) + pnorm(30*(.2 - x2))  ) - 1  )*
      ((  pnorm( 30*(x1 - .0)) + pnorm(30*(.6 - x1))  ) - 1  )*
      (   pnorm( 10*(.2 - x3))  ) +
      
      (  1 - 8*(x1 + x2 - x4 - x5 - x6)^2 )*
      ((  pnorm( 40*(x2 - .0)) + pnorm(40*(.2 - x2))  ) - 1  )*
      ((  pnorm( 40*(x1 - .6)) + pnorm(40*(1 - x1))  ) - 1  )*
      (   pnorm( 10*(.2 - x3))  ) +
      
      .5*(  1 - sin( 8*pi*x1 + 7*pi*x2*x3 - 4*pi*x4*x5*x6) )*
      ((  pnorm( 30*(x2 - .0)) + pnorm(30*(.3 - x2))  ) - 1  )*
      (   pnorm( 8*(x3 - .3))  ) +
      
      (  5*cos( 2*(x2 + .5)*(-x4 + .5)*(-x5+.5)^2 )*(-x6 - .5) - 
           .02*((1-x2)^2 + 
                  (1-x1)^2 + 
                  (1-x3  - .3*x4)^2  + 
                  (1-x5  + .5*x4)^2 + 
                  (.8-x6 - .4*x4)^2 )   )*
      ( pnorm(5*(x2 - 1) + pnorm(10*(.5 - x3))))
    
    return(Y)
    
    
  }# end simba
}# end function

# maximum for Simba
X = matrix( c(.523, .0999, 0,0.298, 0.298, 0.245), ncol= 6)
truth(X, name = "simba", P = 6)

# maximum for Beach
X = matrix( c( 1, 0.85, 1, 0, 0,0), ncol= 6)
truth(X, name = "beach", P = 6)

# maximum for Drum
X = matrix( c( .368, .533, 0, 1, 0.555, 1), ncol= 6)
truth(X, name = "drum", P = 6)
\end{verbatim}

\bibliographystyle{asa}
\bibliography{des}